\documentclass[lettersize,journal]{IEEEtran}
\usepackage{amsmath,amsfonts}
\usepackage{algorithmic}
\usepackage{algorithm}
\usepackage{array}
\usepackage{textcomp}
\usepackage{stfloats}
\usepackage{url}
\usepackage{verbatim}
\usepackage{graphicx}
\usepackage{cite}
\usepackage{subfigure}
\usepackage{enumitem}
\usepackage{caption}
\usepackage{multirow} 
\usepackage{multicol} 
\usepackage{booktabs}
\usepackage{xcolor}
\usepackage{hyperref}

\newcommand{\lyy}[1]{\textcolor{black}{#1}}

\begin{document}

\title{A Survey on Recommendation Unlearning: Fundamentals, Taxonomy, Evaluation, and \\Open Questions}

\author{Anonymous Authors}
\author{Yuyuan~Li,
        Xiaohua~Feng,
        Chaochao~Chen{*},~\IEEEmembership{Senior Member,~IEEE}, 
        and~Qiang~Yang,~\IEEEmembership{Fellow,~IEEE}

\thanks{Y. Li is with the School
of Communication Engineering, Hangzhou Dianzi University, Hangzhou, China.}
\thanks{X. Feng and C. Chen are with the College of Computer Science and Technology, Zhejiang University, Hangzhou, China.}
\thanks{Q. Yang is with Hong Kong Polytechnic University, Academy for AI, Kowloon, Hong Kong, China.}
\thanks{* C. Chen is the corresponding author. Email: {zjuccc@zju.edu.cn.}}}

\markboth{Journal of \LaTeX\ Class Files,~Vol.~14, No.~8, August~2021}%
{Shell \MakeLowercase{\textit{et al.}}: A Sample Article Using IEEEtran.cls for IEEE Journals}


\maketitle

\begin{abstract}
  Recommender systems have become increasingly influential in shaping user behavior and decision-making, highlighting their growing impact in various domains.
  Meanwhile, the widespread adoption of machine learning models in recommender systems has raised significant concerns regarding user privacy and security. 
  As compliance with privacy regulations becomes more critical, there is a pressing need to address the issue of recommendation unlearning, i.e., eliminating the memory of specific training data from the learned recommendation models.
  Despite its importance, traditional machine unlearning methods are ill-suited for recommendation unlearning due to the unique challenges posed by collaborative interactions and model parameters.
  This survey offers a comprehensive review of the latest advancements in recommendation unlearning, exploring the design principles, challenges, and methodologies associated with this emerging field.
  We provide a unified taxonomy that categorizes different recommendation unlearning approaches, followed by a summary of widely used benchmarks and metrics for evaluation.
  By reviewing the current state of research, this survey aims to guide the development of more efficient, scalable, and robust recommendation unlearning techniques. 
  Furthermore, we identify open research questions in this field, which could pave the way for future innovations not only in recommendation unlearning but also in a broader range of unlearning tasks across different machine learning applications.
\end{abstract}


\section{Introduction}

\IEEEPARstart{O}{ver} the past few years, rapid advancements have propelled machine learning, particularly deep learning, to unprecedented heights~\cite{kingma2014adam, lecun2015deep, he2016deep, vaswani2017attention, ho2020denoising, radford2019language, NEURIPS2020_1457c0d6, openai2023gpt}.
Innovations in algorithms, architectures, and computing power have enabled models to achieve remarkable performance across diverse tasks, revolutionizing industries and reshaping societal norms~\cite{silver2016mastering, silver2017mastering, jumper2021highly, biderman2023pythia, wu2023bloomberggpt, touvron2023llama, dubey2024llama, abramson2024accurate, dettmers2024qlora}.
Concurrently, exponential growth in data usage for training machine learning models has brought forth pressing concerns, particularly about privacy and security~\cite{stahl2018ethics, eshete2021making, zhang2024instruction}.
As more and more personal, sensitive, and potentially harmful data is leveraged for model training, concerns about how this data is collected, stored, and used have intensified. Individuals now face increasing risks to their privacy, with potential misuse or unauthorized access to personal information becoming more prevalent. 
In this context, regulatory measures such as the General Data Protection Regulation~\cite{euro2018gdpr}, the California Consumer Privacy Act~\cite{cal2018ccpa}, and the Delete Act~\cite{cal2023del} now empower individuals to exercise their \textit{right to be forgotten}, demanding the removal of personal data used during model construction.

The challenge, however, extends beyond the simple deletion of data.
Machine learning models have been shown to exhibit the ability to memorize or retain information from their training data~\cite{jia2018attriguard, hu2022membership}.
This means that even if specific data points are removed from a dataset, traces of that data might persist within the model itself, potentially leading to privacy breaches or non-compliance with regulatory requests. 
This issue has spurred the development of \textit{machine unlearning}, i.e., a novel field of research aimed at effectively forgetting specific data that was used during model training~\cite{bourtoule2021machine, wang2024machine, liu2024threats, shaik2024exploring,liu2024machine, li2024machine, liu2024survey, liu2024survey2}. 
While machine unlearning was initially driven by the need to comply with emerging regulations, its scope has expanded beyond legal compliance, revealing significant potential for improving model quality and performance. Unlearning has proven to be a valuable tool not only for addressing privacy concerns but also for removing data that is inaccurate, toxic, biased, or poisoned~\cite{li2016data, mehrabi2021survey, xing2024efuf}. 
By proactively identifying and eliminating such harmful data, machine unlearning can contribute to the creation of more reliable, fair, and robust machine learning models.

\begin{figure*}[t]
  \centering
  \includegraphics[width=0.9\linewidth]{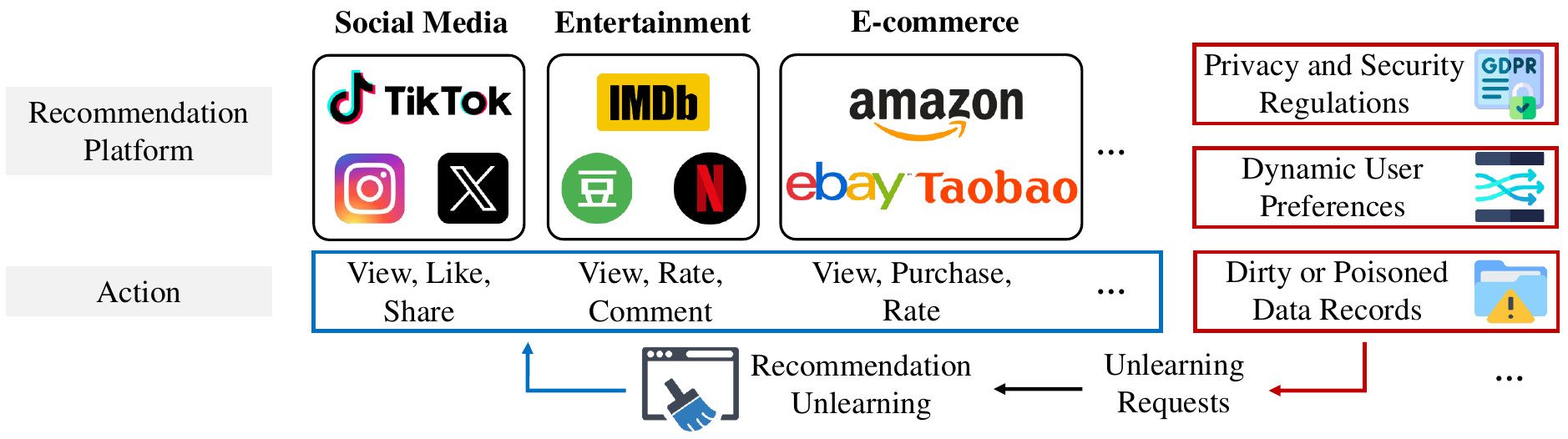}
  \caption{Illustration of application scenarios of recommendation unlearning (e.g., social media, entertainment, and e-commerce).}
  \label{fig:app}
\end{figure*}

Recommender systems are designed to predict user preferences by analyzing historical interactions, which typically include actions such as clicks, purchases, ratings, and likes.
These interactions serve as a foundation for making personalized recommendations and are central to the functioning of many modern digital platforms, from e-commerce websites to streaming services. 
As machine learning models become more deeply integrated into recommender systems~\cite{he2017neural, wang2019neural, chen2020efficient, he2020lightgcn, tan20224sdrug, su2023personalized, acharya2023llm, lin2024data}, the need for recommendation unlearning has gained significant attention as a critical research area~\cite{chen2022recommendation}.
There are two primary factors that drive the growing necessity of recommendation unlearning.
Firstly, historical interactions tend to be intimate and closely tied to user privacy~\cite{zheng2022matrix, liu2023differentially, ijcai2024p0238, liu2024reducing, su2024revisit}, thus carrying a higher likelihood of unlearning requests.
Secondly, the precision of recommendation models critically depends on the quality of training data~\cite{schafer2007collaborative}, mandating rigorous cleansing regimes to remove dirty data that affects the recommendation performance.
As shown in Figure~\ref{fig:app}, from a practical standpoint, recommendation unlearning involves withdrawal actions within recommendation platforms. 
This can include activities such as revoking or modifying past likes, ratings, or other forms of feedback, reflecting the dynamic nature of user engagement with platforms. 
Users may update their preferences, delete or modify past interactions, or request that certain types of recommendations be forgotten. 
The ability to unlearn data from a model ensures that it remains up-to-date, accurate, and in compliance with user requests and regulatory requirements.

However, traditional unlearning methods, primarily tailored for image classification tasks, are poorly suited to the complex requirements of recommendation unlearning~\cite{li2023ultrare}. 
This misalignment arises from the fundamentally different nature of recommender systems, which rely heavily on collaborative interactions between users and items to provide personalized and relevant suggestions. 
In contrast, image classification tasks typically involve a more straightforward relationship between \lyy{images} and labels, without the \lyy{collaborative} dependencies that characterize recommender systems.
Traditional unlearning methods, if applied indiscriminately in recommender systems, risk disrupting these delicate collaborative relationships, leading to the loss of crucial patterns that underpin accurate recommendations~\cite{chen2022recommendation, li2023selective, li2024making}. 
Furthermore, the vast parameters of recommendation models, particularly user and item embeddings, exacerbate the inefficiency issue of traditional unlearning~\cite{li2023selective}.
\lyy{Thus, the collaborative dependencies can create additional challenges for recommendation unlearning in terms of target identification, method design, and unlearning evaluation.}
Given these unique challenges, it is clear that recommendation unlearning cannot rely on traditional unlearning techniques. Instead, new and specialized methods are required.

Given the growing significance of recommendation unlearning, only one existing survey has reviewed the existing techniques employed in this area~\cite{sachdeva2024machine}.
While such a survey \lyy{provides} valuable insights into \lyy{current methods}, it focuses narrowly on the technical aspects, \lyy{failing to establish a systematic taxonomy of these methods.
Moreover, it provides limited information on evaluation methodologies, emerging challenges, and future opportunities in the domain.
This narrow scope leaves significant gaps in our comprehensive understanding of recommendation unlearning.}
In light of this gap, the aim of this survey is to \lyy{deliver} a more comprehensive and in-depth review of \lyy{i) technical approaches along with their taxonomy and design principles, ii) evaluation methodologies and widely-used metrics, iii) summarization of existing methods and research trends, and iv) open questions and future research directions.}
We summarize the main contribution of our survey as follows:
\begin{itemize}
    \item \textbf{Foundational Introduction} (Section~\ref{sec:pre}): Beginning with foundational concepts of recommendation unlearning, we establish a clear understanding of \lyy{the} recommendation unlearning target, unlearning workflow, and design principles that guide the development of unlearning techniques.
    \item \textbf{Taxonomy Development} (Section~\ref{sec:tax}): We present a unified taxonomy of existing recommendation unlearning methods to clarify the unlearning targets, main techniques, and focused problems of existing methods.
    This taxonomy provides clarity on the diversity of approaches and highlights the nuances that distinguish them.
    \item \textbf{Evaluation Summary} (Section~\ref{sec:eva}): We summarize the evaluation resources for recommendation unlearning, including the commonly used datasets, recommendation models, and evaluation metrics regarding each design principle.
    By providing a clear mapping of these resources, we aim to assist researchers in selecting the appropriate tools for evaluating the effectiveness of different unlearning approaches.
    \item \textbf{Open Questions Exploration} (Section~\ref{sec:cha}): We pose the remaining challenges, probe open research questions in the field, aiming to explore the potential research direction and to inspire new approaches for recommendation unlearning.
\end{itemize}

\section{Fundamentals}\label{sec:pre}
In this section, we first introduce the foundational concepts of recommendation unlearning, including the definition of recommendation unlearning, and the basis of recommendation models. 
Then, we delve into the targets and workflow of recommendation unlearning, followed by the design principles of unlearning methods.
While there is a substantial overlap between recommendation unlearning and machine unlearning, we primarily focus on the unique characteristics of recommendation unlearning and provide the related literature for further understanding of machine unlearning.

\begin{figure*}[t]
  \centering
  \includegraphics[width=\linewidth]{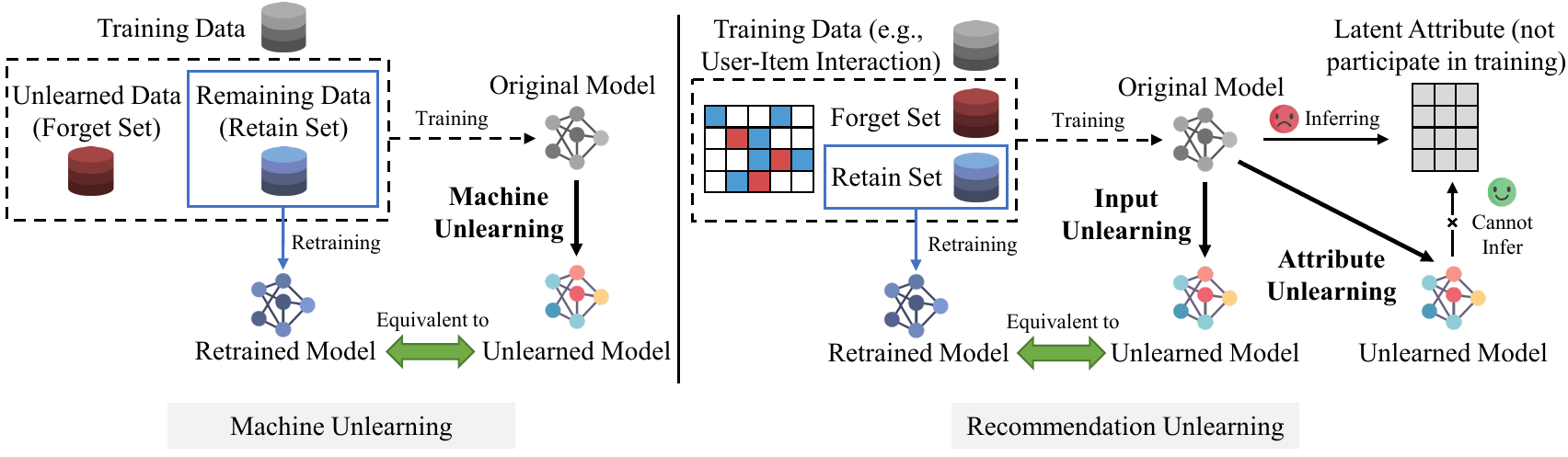}
  \caption{Comparison of machine unlearning and recommendation unlearning.}
  \label{fig:compare}
\end{figure*}

\subsection{Foundational Concepts}\label{sec:concept}

\subsubsection{Machine Unlearning}

Machine unlearning aims to eliminate the memory of specific training from a given model.
Given a dataset $\mathcal{D}$ and a learning algorithm $A(\cdot)$, a model $\theta_0 = A(\mathcal{D})$ can be learned (i.e., the original model before unlearning).
An unlearning task typically specifies a target that needs to be unlearned.
In earlier literature, the unlearning targets were generally confined to the training data, as the model's memory is learned directly from the training data~\cite{cao2015towards, bourtoule2021machine}.
We refer to these traditional unlearning tasks as \textit{input unlearning}, meaning they focus on unlearning the input data used to train the model.
A more detailed discussion of different unlearning targets is provided in Section~\ref{sec:target}.

For clarity and simplicity, \lyy{in input unlearning}, we define the unlearning target here as a subset of the training data (i.e., the forget set $\mathcal{D}_f \in \mathcal{D}$).
The remaining training data is therefore defined as the \lyy{retain} set, where 
$\mathcal{D}_f \cup \mathcal{D}_r = \mathcal{D}$.
As shown in Figure~\ref{fig:compare}, a straightforward method of unlearning is to retrain the model from scratch using the retain set (i.e., $\theta^* = A(\mathcal{D}_r)$)
However, with the increasing size of both data and models in modern applications, retraining from scratch becomes computationally prohibitive and impractical. 
This challenge has, in turn, driven the development of more efficient methods for machine unlearning.
The retrained model $\theta^*$ serves as the ground truth for unlearning.
Therefore, the goal of unlearning is to design an unlearning algorithm $U(\cdot)$ such that 
\begin{gather}
    U(A, \mathcal{D}_f, \mathcal{D}_r, \theta_0) \doteq A(\mathcal{D}_r),\\
    \quad \text{s.t.} \quad Cost( U(A, \mathcal{D}_f, \mathcal{D}_r, \theta_0)) < Cost(A(\mathcal{D}_r)),\nonumber
\end{gather}
where the input of $U(\cdot)$ may be reduced under different conditions (e.g., when the retain set is not available), $Cost(\cdot)$ denotes the computational overhead, and $\doteq$ denotes the degree of unlearning completeness, i.e., to what extent the target has been completely unlearned.
This is also referred to as the effectiveness or efficacy of unlearning in the literature. 
The discussion of evaluation metrics of unlearning completeness will be provided in Section~\ref{sec:prin}.
For clarity, in this paper, we refer to $\theta$ as the original model, $\theta_u = U(A, \mathcal{D}_f, \mathcal{D}_r, \theta_0)$ as the unlearned model, and $\theta^* = A(\mathcal{D}_r)$ as the retrained model.

\lyy{As illustrated in Figure~\ref{fig:compare}, recommendation models may inadvertently encode sensitive attribute information, constituting a distinct category of data requiring protection beyond training datasets.
Crucially, this attribute information is not directly used during model training, meaning that i) it cannot be inferred solely from training data leakage, and
ii) it remains unremovable through retraining from scratch (i.e., all input unlearning methods).
Consequently, attribute unlearning is fundamentally defined as an attack-based paradigm.
A detailed introduction of attribute unlearning and the attack-based evaluation metric are provided in Section III.B and Section IV.B, respectively.}

\subsubsection{Recommender systems} Recommender systems leverage historical interactions to predict user preferences.
\lyy{A fundamental and classical research focus is} the rating prediction task, where users rate various items~\cite{steck2013evaluation}. 
Recommendation models leverage observed (i.e., historical) ratings to predict unobserved ones. 
Collaborative filtering is the foundational approach in modern recommender systems, based on the assumption that users tend to favor similar items, and that an item is likely to be favored by similar users, creating a collaborative effect~\cite{koren2021advances}.
Under this approach, matrix factorization-based models have become widely used in recommender systems, both in academia and industry~\cite{ma2008sorec, hu2008collaborative}.
Existing research on recommendation unlearning also primarily focuses on matrix factorization-based models.
Generally, the core idea behind matrix factorization-based models is to learn a user embedding matrix and an item embedding matrix. The predicted ratings are then obtained through the dot product (or other combination methods) of these two matrices.
Subsequent research has leveraged deep learning techniques~\cite{xue2017deep, he2017neural}, graph learning approaches~\cite{he2020lightgcn, peng2022svd, zheng2021dgcn}, and Large Language Models (LLMs)~\cite{acharya2023llm, lin2024data} to further enhance recommendation performance.

\subsubsection{Recommendation Unlearning} Conducting machine unlearning tasks in recommender systems is referred to as recommendation unlearning. 
In contrast to traditional machine unlearning, where the data record is typically an independent entity such as an image, the data record in recommender systems consists of user-item interactions. Unlearning such a record impacts both user and item, thereby presenting unique challenges of recommendation unlearning.

Recommender systems represent a real-world scenario where unlearning is particularly important. 
\begin{itemize}
    \item First, modern recommender systems widely employ machine learning or deep learning models, which can retain substantial memory of the training data. 
    \item Second, due to the nature of recommendation applications, these systems inherently collect vast amounts of user data, increasing the likelihood of receiving unlearning requests. 
\end{itemize}
For example, a user’s ratings, search queries, or browsing history can reveal much about their preferences, habits, and even personal life. 
As privacy concerns intensify in the digital age, individuals are increasingly seeking control over their data, including the ability to remove or forget data that has been used to train recommender models.
\begin{itemize}
    \item Last but not least, \lyy{due to data sparsity and contextual dependencies among users and items}, recommendation models are highly sensitive to the quality of the training data~\cite{schafer2007collaborative}. This creates a need for systems to unlearn problematic data proactively.
\end{itemize}
In practice, data is often noisy, incomplete, or biased, which can degrade the effectiveness of recommender systems. 
In some cases, data may be deliberately poisoned by malicious users or reflect biased user behaviors, which can negatively affect the system’s outputs. 
To mitigate these risks, recommender systems must undergo regular data cleansing processes to remove or adjust problematic data. 
Recommendation unlearning plays an essential role in this context by providing a way to efficiently remove undesirable data without requiring a full retraining of the model, which can be resource-intensive and time-consuming.

\begin{figure}
  \centering
  \includegraphics[width=\linewidth]{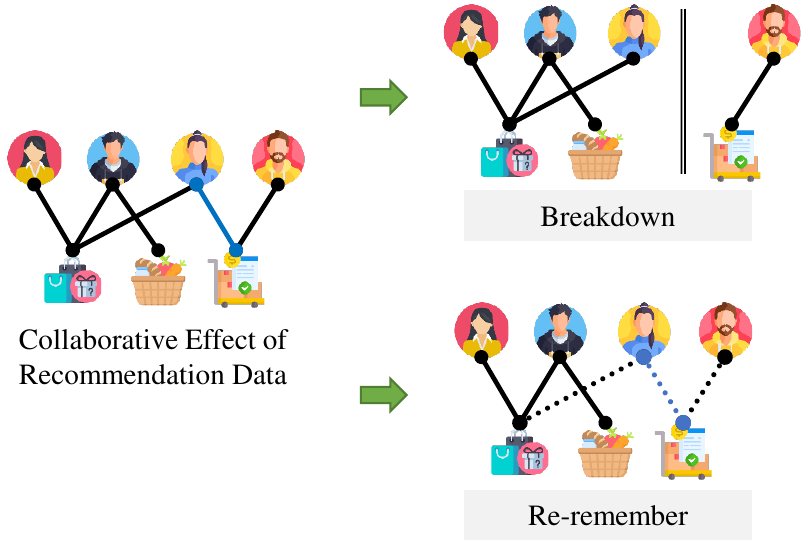}
  \caption{Unique challenges of recommendation unlearning, where black and blue lines represent original and unlearned interactions between users and items, respectively.}
  \label{fig:example}
\end{figure}

However, traditional machine unlearning methods cannot be indiscriminately applied to recommender systems, due to the unique challenges posed by collaborative interactions and parameters of recommendation models.
\lyy{The collaborative nature between user-item interactions introduces contextual dependencies, meaning a change in one data point can propagate through the system, affecting recommendations for multiple users or items.}
Firstly, \lyy{traditional exact unlearning methods often require dividing the dataset.}
As shown in Figure~\ref{fig:example}, \lyy{this additional breakdown of user-item interactions can lead to a decline in recommendation performance.}
Conversely, the retain set can leverage the collaborative effects to re-remember the forget set, thereby causing non-compliance with the unlearning request.
\lyy{Secondly, approximate unlearning methods typically rely on gradient updates, which do not consider the gradients of associated data samples. As shown in Figure~\ref{fig:example}, the collaborative effect may cause the model to re-memorize unlearned data through these associated samples.}
\lyy{Thirdly}, the user and item embeddings, which are crucial parameters in recommendation models, capture the features of users and items. 
As a result, existing recommendation unlearning methods typically focus on manipulating these embeddings. 
However, because the user and item embeddings form high-dimensional and dense matrices, and the number of users and items is massive, applying traditional unlearning methods to these embeddings introduces significant computational overhead, which makes recommendation unlearning even more challenging.

\subsubsection{Unlearning Targets}\label{sec:target}

\begin{figure}
  \centering
  \includegraphics[width=\linewidth]{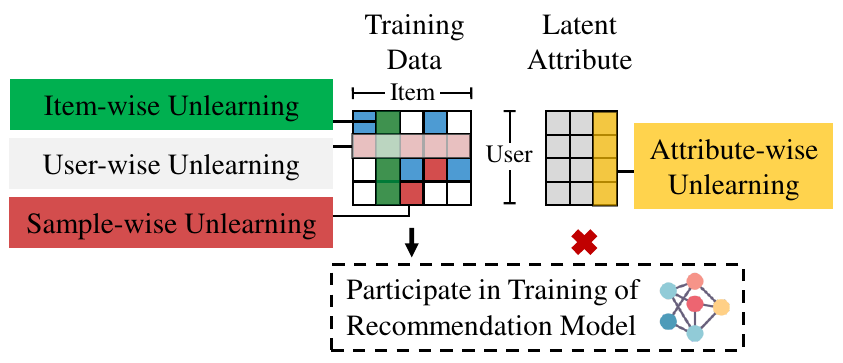}
  \caption{Illustration of different unlearning targets in recommendation unlearning.}
  \label{fig:type}
\end{figure}

Unlearning target is the information that needs to be unlearned, i.e., forget set. 
As mentioned above, most research on unlearning focuses on the task of input unlearning, where the training data used for model training is treated as the unlearning target.
In the context of recommender systems, unlearning targets can be mainly classified into three categories based on the scope of the training data: user-wise, item-wise, and sample-wise.
As shown in Figure~\ref{fig:type}, user/item-wise unlearning targets involve unlearning all samples associated with a specific user/item.
Sample-wise targets are more granular than user/item-wise targets, allowing for the selective unlearning of specific samples.
User-wise and sample-wise unlearning targets are more commonly favored in research, as these targets are believed to contain user-related privacy information, which has a higher likelihood of being unlearned.
In most rating prediction tasks, e.g. a user-item interaction matrix is the training data, user-wise unlearning methods can be directly adapted to item-wise unlearning. This is because, from a matrix perspective, user-wise and item-wise unlearning are essentially equivalent.

In addition to the training data, data that does not participate in training may also need to be unlearned in recommender systems. 
This is because adversaries can potentially infer private information from a trained model, even if that information was never explicitly included in the training data. 
This type of information, referred to as attributes, is implicitly learned by the model during training. 
Such attacks are known as attribute inference attacks~\cite{gong2018attribute, zhao2021feasibility, jayaraman2022attribute}.
The task of unlearning these attributes is called \textit{attribute unlearning}, which serves as a defense mechanism against attribute inference attacks~\cite{ganhor2022unlearning, li2023making, chen2024post}.
As shown in Figure~\ref{fig:compare}, the unlearning target of attribute unlearning is the latent user attributes that are not part of the training data, e.g., gender, age, and race.

\subsubsection{Unlearning Workflow}

\begin{figure}
  \centering
  \includegraphics[width=0.4\textwidth]{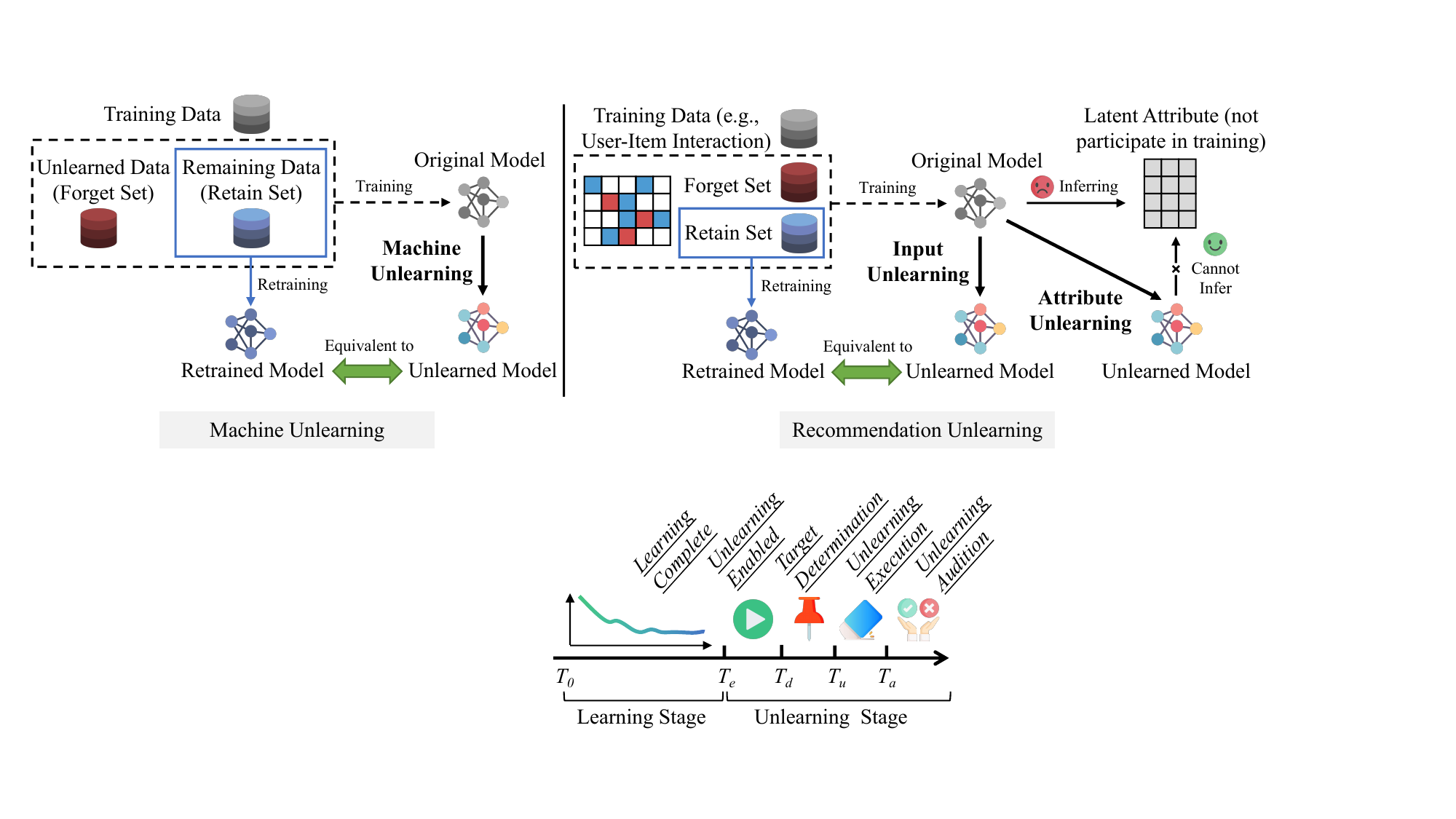}
  \caption{Illustration of unlearning workflow.}
  \label{fig:workflow}
\end{figure}

The unlearning process is conducted under specific conditions. 
As shown in Figure~\ref{fig:workflow}, the unlearning workflow consists of both learning and unlearning stages. 
The learning stage is included because the target of unlearning is the knowledge previously learned by the model, and the model must fully learn this knowledge before unlearning can take place.

The time point $T_e$ marks the separation between the learning and unlearning stages. 
After $T_e$, unlearning becomes enabled.
The unlearning stage itself consists of three steps: i) Target Determination. After $T_d$, the unlearning request is received, and the unlearning target is identified.
ii) Unlearning Execution. After $T_u$, the unlearning process begins. This is the step where most existing studies focus.
iii) Unlearning Audition. This step audits whether unlearning has been successfully executed, i.e., whether the unlearning target has been completely removed from the model's memory.

\subsection{Design Principles}\label{sec:prin}

The goal of unlearning is not merely to eliminate the memory of the target being unlearned. 
It encompasses broader goals.
Generally, there are three key design principles for unlearning methods, which are also applicable to recommendation scenarios~\cite{chen2022recommendation, li2024making, li2023ultrare}.

\paragraph{Unlearning Completeness}\label{sec:comp}
Completely unlearning the memory of target data is one of the most fundamental goals of unlearning.
As discussed in Section~\ref{sec:concept}, the completeness of unlearning is defined as the equivalence between the unlearned model and the retrained model.
However, the definition of this equivalence varies in the literature, leading to different evaluation metrics. 
We categorize the definition of completeness equivalence into the following four perspectives.
\begin{itemize}
    \item Algorithmic perspective: Only retraining from scratch (i.e., unlearning from the algorithmic level) satisfies the definition of complete unlearning in this perspective.
    Thus, this definition of completeness can only be self-evaluated algorithmically or verified by providing training checkpoints~\cite{jia2021proof, thudi2022necessity, weng2024proof}.
    Exact unlearning methods adhere strictly to this definition, designing efficient strategies to achieve retraining. 
    All other unlearning methods are classified as approximate unlearning~\cite{nguyen2022survey}.
    \item Parametric perspective: This perspective defines equivalence at a parametric level, meaning that the goal is to achieve parameters of the unlearned model similar to those of the retrained model. 
    Since the training of machine learning models involves randomization (e.g., initialization seed and batch order), ``similarity'' is typically defined as being close in distribution. 
    Influence function-based unlearning methods tend to favor this definition, as influence functions provide a closed-form approximation of the retrained model.
    \item Functional perspective: This perspective focuses solely on equivalence at a functional level, aiming to ensure that the unlearned model behaves like the retrained model.
    Specifically, this means the unlearned model should perform poorly on the forgot set while maintaining its original performance on the retain set.
    This perspective is often preferred in practice, as the model's output is the critical factor in most real-world applications. 
    The relaxed definition of unlearning completeness also allows for the use of a broader range of techniques.
    \item Attack perspective: In this attack perspective, complete unlearning is defined as making it impossible for adversaries to recover the unlearning target. 
    By leveraging additional information (e.g., when adversaries exploit the difference between the unlearned model and the original model), this definition of attack-level unlearning could challenge the previous definitions, including the algorithmic one. 
    Therefore, this definition also offers an alternative perspective for evaluating completeness.
\end{itemize}
The specific examples of evaluation metrics are introduced in Section~\ref{sec:metric}.
\lyy{From the point of unlearning techniques, these four perspectives form a continuum from strict to relaxed definition: beginning with the algorithmic requirement, progressing to parametric similarity, then to functional equivalence, and finally to resilience against adversarial attacks. 
However, in practical applications, the attack and functional perspectives are particularly significant, as they directly reflect model behavior and security in real-world scenarios.
Recent study has demonstrated that satisfying the algorithmic perspective does not necessarily guarantee completeness under the functional or attack perspectives~\cite{CURE4Rec2024}.}

\paragraph{Unlearning Efficiency}
Given the significant computational cost associated with complex recommender models and large datasets in real-world applications, improving unlearning efficiency, particularly in terms of time, is a crucial goal of unlearning.

\paragraph{Model Utility}
The performance of a model, i.e., model utility, relies on the knowledge learned from the training data. 
Removing too much training data inevitably undermines the learned knowledge, which in turn diminishes model utility. 
However, an adequate unlearning method should be able to achieve performance that is comparable to that of a retrained model.
Avoiding further reduction of model utility (i.e., over-unlearning) is another important goal of unlearning.

\begin{figure*}[t]
  \centering
  \includegraphics[width=0.8\linewidth]{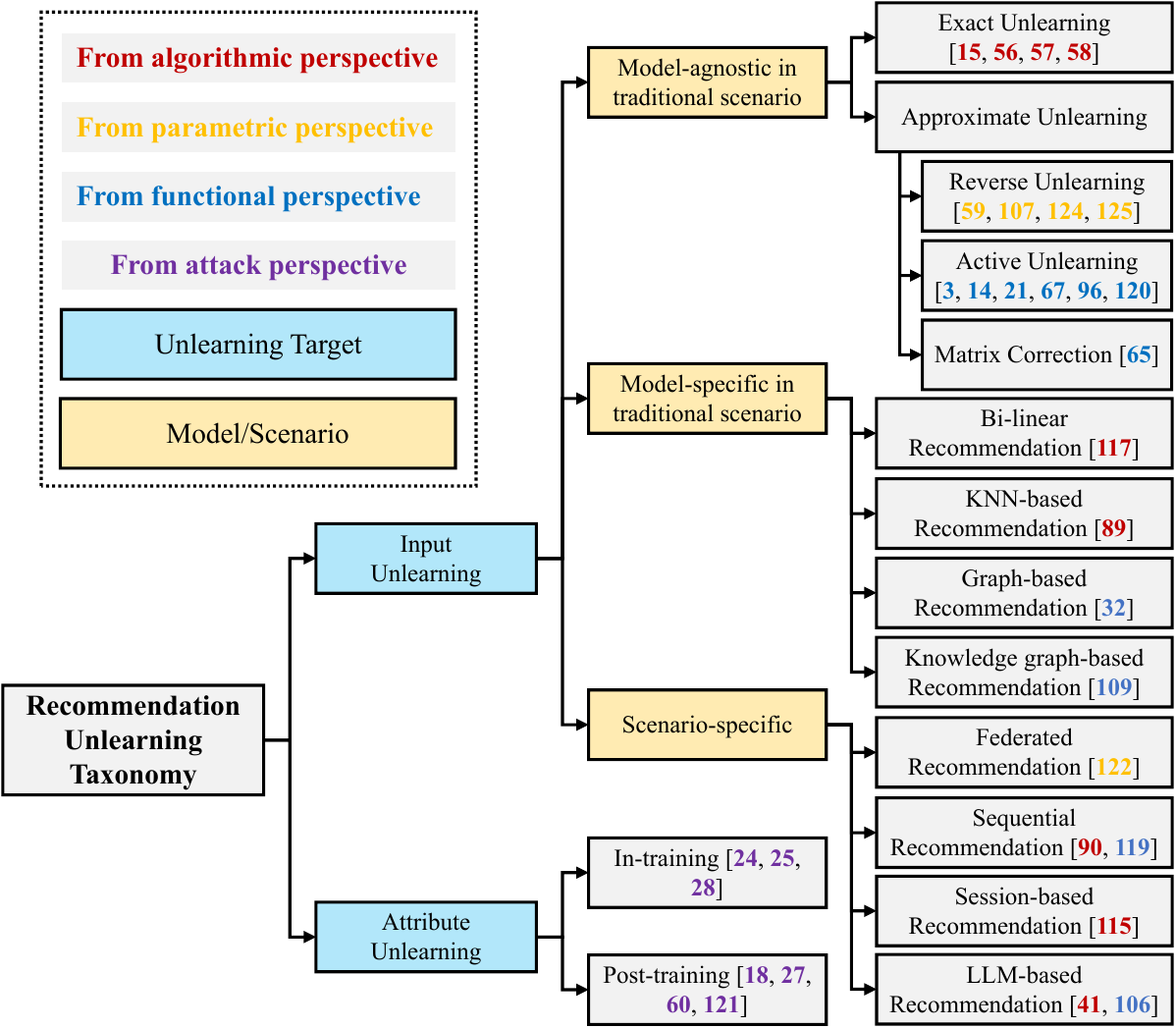}
  \caption{Taxonomy of recommendation unlearning.}
  \label{fig:tax}
\end{figure*}

\section{Taxonomy}\label{sec:tax}

In this section, we present a unified taxonomy that organizes the existing recommendation unlearning methods, providing a clear and structured overview of the field. 
As shown in Figure~\ref{fig:tax}, we first categorize these methods based on the unlearning target into two primary categories: input unlearning and attribute unlearning.
By distinguishing between these two primary categories, we lay the groundwork for a more detailed exploration of specific methods within each category, highlighting the various techniques that have been developed to address the unique challenges of recommendation unlearning.
Additionally, we present key takeaways in Table~\ref{tab:all}, which summarizes the focused problem, the main techniques employed, and the applicable unlearning targets (e.g., user/item/sample/attribute).

\subsection{Input Unlearning}

\lyy{In traditional recommendation scenarios}, based on the types of models that can be applied, existing input unlearning methods can be categorized into two main approaches: model-agnostic and model-specific. 
The term \textit{model-agnostic} refers to methods that can be applied regardless of the model structure, while \textit{model-specific} refers to methods designed for a particular type of model. 
\lyy{In addition to the traditional scenario}, we also review the unlearning methods for recommendation tasks in specific scenarios, such as federated learning and LLMs, where the recommendation model differs from traditional recommender systems.

\subsubsection{Model-agnostic Methods}

Model-agnostic unlearning methods for recommendation tasks are largely inspired by traditional machine unlearning techniques used in classification tasks. Based on the definition of unlearning completeness, we further classify the model-agnostic approach into exact unlearning and approximate unlearning~\cite{nguyen2022survey}.

\begin{figure}[t]
  \centering
  \includegraphics[width=\linewidth]{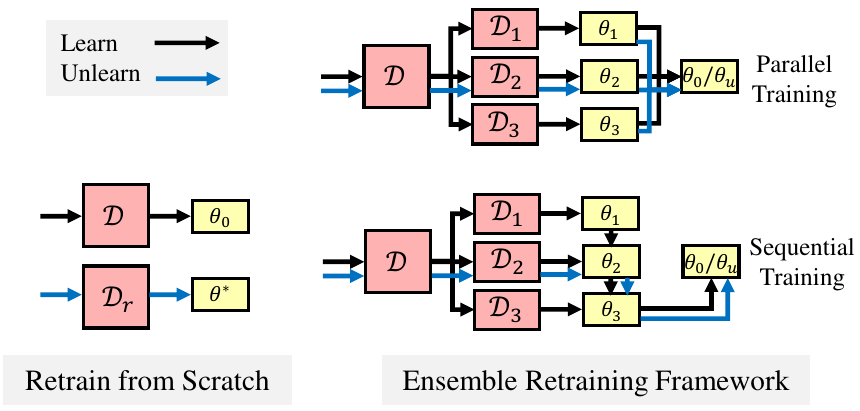}
  \caption{Different designs for exact unlearning in recommender systems, where \lyy{$\theta_0$,  $\theta^*$, and $\theta_u$ denote the original model, the model retrained from scratch, and the unlearned model respectively, $\mathcal{D}_i$ represents the subset of original dataset $\mathcal{D}$, and $\theta_i$ represents the corresponding model trained on $i$-th subset.} The black and blue arrows depict the workflows of learning and unlearning respectively.}
  \label{fig:exact}
\end{figure}

\paragraph{Exact Unlearning} As mentioned in Section~\ref{sec:comp}, exact unlearning follows a strict definition of unlearning completeness, achieving it at the algorithmic level.
Inspired by SISA~\cite{bourtoule2021machine}, exact recommendation unlearning methods predominantly adopt the ensemble retraining framework. 
As shown in Figure~\ref{fig:exact}, this framework divides the original dataset into multiple subsets, trains a sub-model on each subset, and aggregates all sub-models into the final model, similar to an ensemble learning system. 
Note that, to ensure algorithmic unlearning completeness, the design of sub-models is typically identical to that of the original model, including the model structure, hyperparameter settings, and other factors.
This framework allows for more efficient retraining during unlearning. When an unlearning request is submitted, the ensemble retraining framework only needs to retrain the subset containing the unlearning targets, avoiding the need to retrain the entire dataset from scratch, thereby enhancing unlearning efficiency.

Li et al.~\cite{li2024enhancing} directly apply SISA to recommendation models in intelligence education.
Their approach enhances the personalization and accuracy of educational recommendations by selectively forgetting the data inputs for each user.

Building on the design of SISA, Chen et al. propose RecEraser, which introduces two key modifications tailored for recommendation tasks~\cite{chen2022recommendation}.
First, RecEraser incorporates a balanced clustering module for dataset division, grouping similar users or items into the same subset to preserve the collaborative effects within the data, in contrast to the random division used in SISA. 
Second, RecEraser adds an attention network to learn the weights for the weighted aggregation of sub-models. 
This adaptive weighted aggregation, compared to the average weighting or majority voting in SISA, further enhances recommendation performance for the ensemble retraining framework.

Due to the collaborative effect of recommendation data, there is a significant trade-off between unlearning efficiency and model utility. 
Specifically, increasing the number of dataset divisions can enhance unlearning efficiency, but this also disrupts the collaboration among data, which in turn reduces model utility. 
To address this issue, Li et al. propose UltraRE, a lightweight modification of RecEraser~\cite{li2023ultrare}. 
UltraRE introduces a new balanced clustering algorithm based on optimal transport theory, which improves both efficiency and clustering performance simultaneously. Additionally, UltraRE simplifies the attention network used during aggregation, replacing it with a logistic regression model to further enhance efficiency.

To further enhance model utility, LASER adopts sequential training during aggregation, rather than parallel training~\cite{li2024making}. As shown in Figure~\ref{fig:exact}, sequential training involves training one model on a data subset sequentially. This approach helps mitigate the negative impact of dataset division on collaboration.
LASER introduces the concept of curriculum learning to optimize the training sequence of data subsets, thereby improving model utility.
However, sequential training also reduces unlearning efficiency. To address this issue, LASER introduces early stopping and parameter manipulation.

However, there are several drawbacks of \lyy{model agnostic} exact unlearning. 
\lyy{Note that not all exact unlearning methods rely on ensemble retraining frameworks, nor do they all suffer from the following drawbacks. Some model-specific approaches can achieve high efficiency, e.g., KNN unlearning~\cite{schelter2023forget}.}
\begin{itemize}
    \item Exact unlearning requires reformulating the learning process, meaning it cannot be directly applied to an already trained model, which creates significant inconvenience in practical implementation.
    \item \lyy{When samples from the forget set are distributed across multiple subsets, all affected subsets require updates. This creates substantial challenges for subset division design.}
    \item The efficiency gains from dataset division are limited and incremental, whereas approximate unlearning methods can often provide efficiency improvements that are orders of magnitude better. Additionally, as noted by~\cite{li2023ultrare}, exact unlearning introduces a trade-off between efficiency and utility, which is particularly significant in recommendation tasks.
    \item \lyy{The core idea of the ensemble retraining framework is trading space for time. Although this framework enhances efficiency, it requires maintaining multiple copies of model parameters, which introduces significant storage overhead. For large-scale recommendation models in practical applications, this creates both storage capacity challenges and potential security concerns.}
    
    \item The performance of exact unlearning is suboptimal in practice~\cite{CURE4Rec2024}. Although exact unlearning achieves perfect completeness in theory, its empirical performance is unsatisfactory, often worse than approximate unlearning. In real-world scenarios, users prioritize good performance over theoretical guarantees.
\end{itemize}

\begin{figure}[t]
  \centering
  \includegraphics[width=\linewidth]{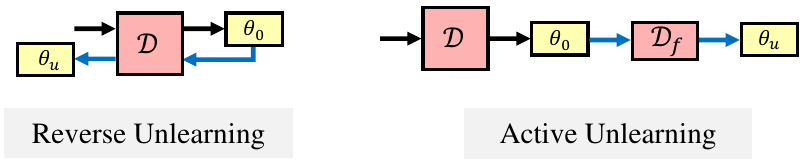}
  \caption{Different designs for approximate unlearning in recommender systems, i.e., reverse unlearning and active unlearning. \lyy{$\theta_0$ and $\theta_u$ denote the original model and the unlearned model respectively. $\mathcal{D}$ and $\mathcal{D}_f$ represents the original datasets and the forget set respectively.}}
  \label{fig:appro}
\end{figure}

\paragraph{Approximate Unlearning} Apart from adopting retraining as exact unlearning, all other unlearning methods are classified as approximate unlearning. These methods either approximate the unlearned model from a parametric perspective or from a functional perspective.
As shown in Figure~\ref{fig:appro}, reverse unlearning estimates the influence of the unlearning target and directly obtains the unlearned model without additional training, thereby approximating from a parametric perspective. In contrast, active unlearning fine-tunes the model to obtain the unlearned model, approximating from a functional perspective.

\textbf{Reverse Unlearning.} Deep learning models are typically trained using gradient descent-based optimization. An intuitive approach for unlearning is to add back the gradient of the target data that was previously subtracted, thereby mitigating its influence on the model and achieving the goal of unlearning from a parametric perspective. 
Formally, let us assume that $z_i \in \mathcal{D}$ is a data record from the training set, and the recommendation loss function is $\ell(z_i, \theta)$.
We further assume that the loss function is twice-differentiable and strictly convex, which holds true for the majority of recommendation loss functions.
\lyy{Note that these assumptions serve only as foundational simplifications for reduction-based analyses and do not imply convexity in deep learning-based recommendation models.}
Following the empirical risk minimization framework, the optimal model parameter  is given by
\begin{equation}
    \theta_0 = \underset{\theta}{\operatorname{\arg\min}}\hspace{1mm}\sum_{i=1}^n \ell(z_i, \theta),
\end{equation}
where $n$ is the number of records, and for simplicity, we omit the regularization term in the loss function.
According to the definition of algorithmic-level unlearning completeness, unlearning requires generating a model that retrains from scratch without the target data record.
Thus, unlearning a data point $z$ can be formulated as:
\begin{equation}\label{equ:rev}
    \theta^* = \underset{\theta}{\operatorname{\arg\min}}\hspace{1mm}\sum_{i=1}^n \ell(z_i, \theta) - \ell(z, \theta).
\end{equation}

The early study in traditional machine unlearning explores directly adding back the first-order gradient~\cite{graves2021amnesiac}. 
However, due to the stochastic nature and randomness inherent in deep learning, the first-order gradient is not always accurate for estimating the influence of the target data.
Consequently, subsequent research has focused on more precisely estimating the influence and directly subtracting this influence from the original model~\cite{sekhari2021remember}. This process is simply formulated as:
\begin{equation}
    \theta^{\lnot z}_{u} = \theta - \mathcal{I}(z),
\end{equation}
where $\theta^{\lnot z}_{u}$ is the model that is unlearned with data record $z$, and $\mathcal{I}(z)$ denotes the estimated influence of $z$.
Note that this formulation can be easily generalized to unlearning a set of data records. For the sake of simplicity and clarity, we use a single data record as an example.
Compared to exact unlearning, the main advantage of reverse unlearning is its ease of implementation. It only requires direct manipulation of the model parameters, without interfering with the original training workflow, and can be applied to an already trained model.

Current reverse unlearning methods in recommender systems mainly rely on influence function~\cite{koh2017understanding, koh2019accuracy, basu2020second} to estimate the influence of target data on model parameters.
Specifically, it estimates the influence by weighting a data record by $\epsilon$.
Formally, the $\epsilon$-weighted model parameter is
\begin{equation}
    \theta_{\epsilon, z} = \arg\min_{\theta}\sum_{i=1}^n \ell(z_i, \theta) + \epsilon\ell(z, \theta).
\end{equation}
According to~\cite{koh2017understanding}, leverage second-order \lyy{Taylor} expansion, the estimated influence of $z$ is given by
\begin{equation}\label{equ:if}
    \mathcal{I}(z) := \frac{d\theta_{\epsilon, z}}{d\epsilon}\Big\vert_{\epsilon=0} = -H_{\theta_0}^{-1}\nabla_\theta \ell(z, \theta_0),
\end{equation}
where $\nabla_\theta\ell(z, \theta_0)$ is the gradient vector, and $H_{\theta_0} := \sum_{i=1}^n\nabla^2_\theta\ell(z, \theta_0)$ is the Hessian matrix and is positive definite by assumption.
The derivation in Eq.(\ref{equ:if}) is equivalent to a single step of the Newton optimization update. Therefore, influence function-based reverse unlearning methods can be interpreted as performing a one-step reverse Newton update.
Zhang et al.~\cite{zhang2023closed} directly apply this closed-form one-step reverse Newton update to the matrix factorization model.
There are also other tasks in recommender systems that utilize the influence function. For example, Wang et al.~\cite{wang2024would} study the user-controllable recommendation task, where users can control which interactions are used for training. This can also be formulated as a recommendation unlearning task.

Although influence function-based unlearning methods theoretically provide a promising solution for recommendation unlearning, they face significant challenges in terms of computational efficiency and estimation accuracy.

Firstly, due to the large number of users and items in recommendation tasks, the user and item embeddings form high-dimensional dense matrices, which incur considerable computational overhead when explicitly calculating the influence function. As shown in Eq.(\ref{equ:if}), this calculation involves computing the inverse of the Hessian matrix and then multiplying it by the gradient vector, which has a computational complexity of $O(n^2)$.
Since the forget set typically accounts for a small portion of the training data, Li et al.~\cite{li2023selective} found that the parameter changes before and after unlearning primarily affect the embeddings of the unlearning target. Changes to other parameters (e.g., non-targeted embeddings and parameters unrelated to embeddings) are negligible. Consequently, they propose selectively calculating the influence function for the target embeddings only, thereby reducing computational overhead at the root level.
Similarly, Zhang et al.~\cite{zhang2023recommendation} propose an importance-based pruning strategy to reduce the model size, thereby lowering the overall computational overhead of the influence function. The importance is estimated based on the neighborhood relationships among the unlearning targets.

Secondly, influence function is essentially an approximation, and its accuracy is not always guaranteed, particularly for deep learning models~\cite{basu2021influence}.
In recommendation tasks, removing the influence of target data can have a cascading effect on neighboring data, thereby further compromising model utility. To address this issue, Li et al.~\cite{li2023selective} incorporate a collaborative term into Eq. (\ref{equ:rev}) to mitigate this negative effect. Instead of directly removing the target data, this collaborative term replaces it with the average rating of the neighboring data, which compensates for the inaccurate estimation of influence function.

Nevertheless, there are several drawbacks of reverse unlearning.
\begin{itemize}
    \item There is a trade-off between accurate estimation and computational efficiency. On the one hand, explicitly calculating the influence function incurs considerable computational overhead. On the other hand, accelerating the calculation through approximation inevitably reduces the accuracy of the influence estimation.
    \item The derivation of the influence function assumes that only a small portion of the training data is removed. The tolerance for the volume of the forget set is not comparable to exact unlearning. Moreover, in real-world scenarios where unlearning requests are submitted sequentially and must be processed immediately (i.e., sequential/stream unlearning~\cite{gupta2021adaptive}), the assumption that only a small portion of the data is unlearned no longer holds.
\end{itemize}

\textbf{Active Unlearning.} From a functional perspective, the goal of unlearning is to make the unlearned model perform as if it were trained from scratch. One straightforward approach is to continue the learning process but with a new objective. Active unlearning fine-tunes the model to achieve this, essentially \textit{learning to unlearn}. As a result, the key challenge is designing an appropriate loss function for the fine-tuning process.

Lie et al.~\cite{liu2022forgetting} first propose active unlearning in recommendation tasks by fine-tuning the original model on the retain set, which is faster than retraining the model from scratch. Furthermore, they apply a 2nd-order gradient optimization technique to further enhance the efficiency of this unlearning process.

Alshehri et al.~\cite{alshehri2023forgetting} propose active unlearning by flipping the labels of data in forget set. 
However, fine-tuning the model with label-flipped records may unintentionally overwrite the parameters of the original model, negatively impacting model utility for other users. To address this issue, they also sample data from the retain set during fine-tuning.

Label flipping is limited to binary ratings $\{0, 1\}$, which presents challenges when dealing with value-based ratings. 
To overcome this issue, Sinha et al.~\cite{sinha2024multi} propose flipping the loss function instead of the labels, reversing the direction of the loss (i.e., changing addition to subtraction). They also utilize data from the retain set to prevent over-unlearning. These two loss functions (i.e., flipped loss with the forget set and original loss with the retain set) are linearly combined with a balancing coefficient.
Similarly, You et al.~\cite{you2024rrl} also flip the loss function and tune it on the forget set. The key difference from previous methods is that they solely rely on the flipped loss, updating it in a single step using the Fisher information matrix. This approach does not depend on the retain set, making it more convenient and efficient for implementation in real-world applications.
Dang et al.~\cite{dangefficient} fine-tune the slipped loss by introducing an additional module to enhance utility.

In addition to fine-tuning, Chaturvedi et al.~\cite{chaturvedi2024unlearning} propose a model fusion approach to achieve active unlearning. Specifically, they fine-tune one model using noisy data and another, smaller model using the retained data set. They then apply a convolutional fusion function to combine the models, thereby efficiently enabling unlearning.

However, there are also several drawbacks of active unlearning:
\begin{itemize}
    \item Active unlearning is a functional approach, meaning there is no theoretical guarantee for the obtained results. The termination point of fine-tuning is solely dependent on the model's performance. Although loss flipping can expand the range of applicable ratings, it essentially performs gradient ascent, which is unbounded and can lead to instability during training.
    \item The one-step Fisher information matrix update is theoretically more promising, but it faces the computational challenge of calculating the Hessian matrix. While this can be computed offline, the high memory storage requirements present a significant challenge for many unlearning executors.
\end{itemize}

Another emerging direction for achieving unlearning also takes a functional perspective, but with an approach distinct from ``learning to unlearn''.
Liu et al.~\cite{liu2023recommendation} reveal that collaborative filtering can be formulated as a mapping-based approach, where the recommendations are obtained by multiplying the user-item interaction matrix with a mapping matrix. This formulation simplifies unlearning to the manipulation of the mapping matrix. While this method provides valuable insights into model-agnostic recommendation unlearning, it has some limitations. The arbitrary approximation of recommendation models as mapping-based approaches lacks a solid theoretical foundation, making it highly dependent on the accuracy of the mapping matrix approximation.

\subsubsection{Model-specific Methods}
In addition to model-agnostic methods, which are designed to work with a wide range of recommendation models (typically collaborative filtering), there are also methods specifically tailored to the structure and characteristics of particular model types. These specialized techniques are often more efficient and effective because they leverage the inherent properties of the model architecture.

\begin{itemize}
    \item Bi-linear recommendation model: 
    Xu et al.~\cite{xu2023netflix} propose an exact unlearning method for bilinear recommendation models, which utilize alternating least squares for optimization~\cite{he2016fast}. The core idea of their approach is fine-tuning. In these models, the confidence matrix, which is multiplied by the predicted ratings, plays a key role. By setting the target elements of this matrix to zero during fine-tuning, the method effectively performs exact unlearning. However, the authors also note that exact unlearning may not always be feasible in real-world applications. This is because additional optimization techniques, such as early stopping, are often employed, which can introduce complexities that prevent exact unlearning from being fully realized.
    \item KNN-based recommendation model: $k$-Nearest Neighbor (KNN) is widely used in a variety of recommendation scenarios~\cite{resnick1994grouplens, ludewig2021empirical, ariannezhad2023personalized} due to its several key advantages. These include its transparency and explainability, cost-effectiveness in scaling to industrial workloads, and significantly lower training time. KNN's simplicity and efficiency make it an attractive choice for both research and practical applications in the field of recommender systems.
    Unlike the models discussed in the review, KNN is a non-parametric model, which inherently facilitates completeness by simply removing the unlearning target. Schelter et al.~\cite{schelter2023forget} propose an efficient indexing method to accelerate this unlearning process, making it both faster and more scalable.
    \item Graph-based recommendation model: Graph Neural Network (GNN)-based recommendation models have gained prominence in recent years~\cite{he2020lightgcn, peng2022svd, zheng2021dgcn}. Hao et al.~\cite{hao2024general} propose inserting a learnable delete operator at each layer of the GNN and fine-tuning the model with a linear combination of two objectives. The first component focuses on unlearning the target data, while the second component aims to maintain consistent feature representation, thereby preserving the model's utility.
    \item Knowledge graph-based recommendation model: Knowledge graph-based recommendation models utilize domain-specific knowledge to produce recommendations, making them a crucial category of systems, especially for tackling the cold-start problem. These models apply rules, reasoning, and constraints derived from domain knowledge~\cite{le2023constraint}.
    In a knowledge graph, a record is represented as a triple $<s, p, o>$ where $s$, $p$, and $o$ denote the subject, predicate, and object, respectively. 
    Wang et al.~\cite{wang2024forgetting} propose two types of unlearning for knowledge graph-based recommendation models: (i) passive forgetting, which unlearns the target data based on user requests, and (ii) intentional forgetting, which optimizes the entire knowledge graph. Their unlearning is achieved through rule replacement.

\end{itemize}

\subsubsection{Scenario-specific Methods}
model-specific methods focus on different models, yet remain centered on traditional recommendation tasks. However, beyond the traditional approaches, there are also unlearning methods specifically designed for other, more complex recommendation scenarios.
\begin{itemize}
    \item Federated Recommendation: The federated learning paradigm enables clients to collaboratively train a recommendation model without sharing their original data. 
    Although federated learning offers a safeguard for user privacy by preventing data sharing, it cannot achieve unlearning. 
    In federated recommendation models, each user typically acts as an individual client. Yuan et al.~\cite{yuan2023federated} propose an unlearning method tailored for federated recommender systems. Building on the design of FedEraser~\cite{liu2021federaser}, their approach efficiently recalculates the update gradients for the retain clients by conducting a few-step fine-tuning process to determine gradient direction, which is then scaled by the original update length.
    \item Sequential Recommendation: Sequential recommender systems, which leverage clients' sequential product browsing history, have become indispensable for providing personalized product recommendations. These systems track the order in which users interact with products, allowing them to predict future preferences and suggest relevant items~\cite{quadrana2018sequence}. 
    Ye et al.~\cite{ye2023sequence} build upon the idea of active unlearning by introducing noise into the interaction sequences. They then fine-tune the original model using these noisy sequences to effectively achieve unlearning.
    \lyy{Schelter et al.~\cite{schelter2024snarcase} propose a low-latency unlearning method for next-basket recommendation via incremental view maintenance.}
    \item Session-based Recommendation: Session-based recommendation models have proven effective in predicting users' future interests by leveraging their recent sequential interactions. While these models also emphasize sequence-aware data, the key distinction of session-based approaches lies in their exclusive focus on the current user session~\cite{wu2019session}. Typically, these methods do not rely on long-term user information, tailoring recommendations solely based on the immediate session context.
    Xin et al.~\cite{xin2024effectiveness} focus on unlearning an item from the current session. They follow the idea of exact unlearning and implement an ensemble retraining framework. Similar to the approach in~\cite{chen2022recommendation}, they use balanced clustering for session division and apply an attention network for aggregating sub-models.
    \item LLM-based Recommendation: LLMs, with their vast number of parameters and training on extensive datasets, exhibit remarkable performance across a wide range of tasks~\cite{openai2023gpt, touvron2023llama, dubey2024llama}. The impressive advancements in LLMs have also sparked interest in using them as recommender systems. The effectiveness of LLM-based recommender systems lies in the models' extensive open-world knowledge and their reasoning capabilities~\cite{acharya2023llm, lin2024data}. 
    LLM-based recommender systems acquire recommendation capabilities through instruction tuning on user interaction data. Note that due to the high cost of retraining LLMs, the efficiency of LLM-based recommendation unlearning has become a significant concern.
    To address this issue, Wang et al.~\cite{wang2025towards} leverage Low-Rank Adaptation (LoRA) for lightweight fine-tuning. Building on the idea of active unlearning, and following a similar approach to~\cite{chundawat2023can}, they adopt a teacher-student framework. In this framework, a \textit{remembering} teacher preserves knowledge from the retain set, while a \textit{forgetting} teacher is used to unlearn information from the forget set. Knowledge distillation is then performed using a Kullback-Leibler divergence loss.
    Hu et al.\cite{hu2024exact} also leverage LoRA for fine-tuning, but they follow the idea of exact unlearning (i.e., ensemble retraining framework). Rather than training separate sub-models, which is computationally prohibitive for LLM-based recommendation models, they train an adapter (LoRA module) for each data subset. This approach enables exact unlearning in LLM-based recommender systems, allowing for highly efficiency retraining.
\end{itemize}

\subsection{Attribute Unlearning}

\begin{figure}
  \centering
  \includegraphics[width=\linewidth]{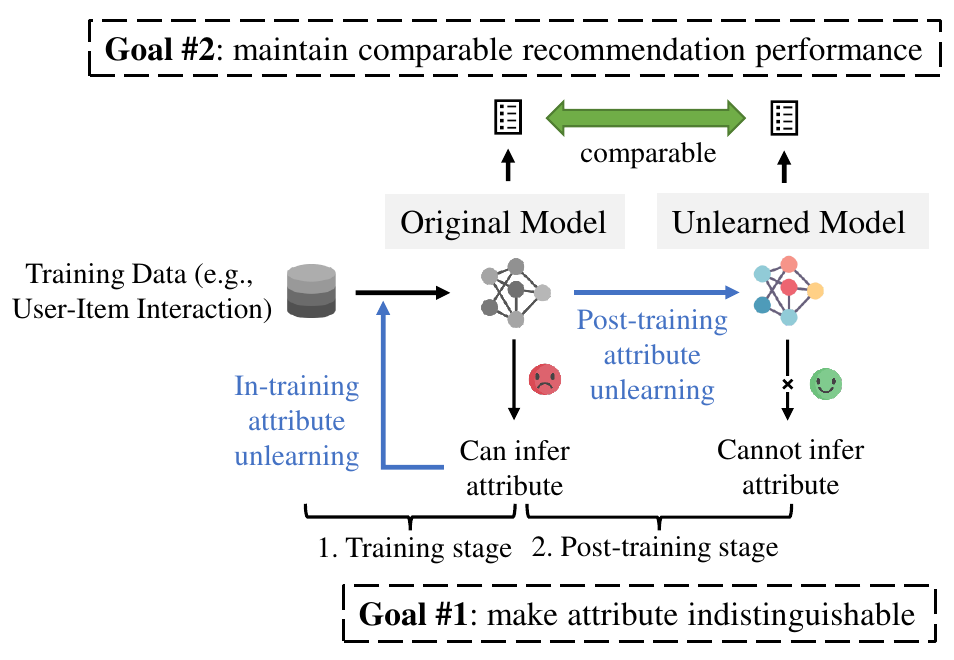}
  \caption{Illustration of in-training and post-training attribute unlearning. Both types of attribute unlearning have the same goals, but they differ in the stage at which the unlearning is executed. Additionally, in-training unlearning can leverage the training data, whereas post-training unlearning typically cannot.}
  \label{fig:attr}
\end{figure}

Due to the powerful modeling capabilities of recommender systems, the embeddings learned by these models can reveal attribute information that was not explicitly encountered during training. Therefore, in addition to input unlearning, there is also a need to unlearn attribute information. 
In general, there are two main goals for attribute unlearning. The primary goal is to unlearn the target attribute information. Since this information is revealed by attribute inference attacks, achieving complete unlearning refers to rendering the attribute indistinguishable to such attackers. 
The second goal is to preserve the recommendation performance. The essential aim of recommender systems is to provide personalized recommendations to users; if the system fails to provide desired services, preserving user privacy through attribute unlearning becomes meaningless.
As shown in Figure~\ref{fig:attr}, based on the stage of execution, attribute unlearning can be categorized into in-training unlearning and post-training unlearning.

\subsubsection{In-training Unlearning} this category performs unlearning during the model's training stage. As a result, in-training attribute unlearning can naturally access the training data, which theoretically helps preserve model utility (i.e., recommendation performance).

Attribute unlearning was first introduced by~\cite{ganhor2022unlearning}, focusing on unlearning attributes in MultVAE. The core idea is adversarial training, which can be extended to other recommendation models as well. Specifically, they incorporated an adversarial decoder into MultVAE, which acts as an attacker attempting to infer user attributes. The encoder, in turn, works to deceive the adversarial decoder through bi-level optimization. As a result, the model learns embeddings that are robust to attribute inference attacks while still effectively modeling user preferences.

Escobedo et al.~\cite{escobedo2024making} propose an alternative approach to in-training attribute unlearning through stereotypicality-based obfuscation. Specifically, they manipulate interaction records by imputation, removal, and weighting, all guided by a stereotypicality score. Interaction records with a high stereotypicality score indicate a strong association with certain attribute information. For example, a male user who predominantly listens to music tracks associated with male stereotypes. By obfuscating these records, they aim to eliminate attribute information while minimizing any negative impact on model utility.

There are two main drawbacks of in-training attribute unlearning.
\begin{itemize}
    \item In-training attribute unlearning requires interference with the model's training process. Similar to exact unlearning, This interference with the training pipeline may lead to delays and operational inefficiencies, especially in large-scale systems, making it less practical for dynamic, real-time environments.
    \item In-training unlearning approach could suffer from Unsatisfactory preservation of model utility. While using training data theoretically improves performance, introducing an unlearning term during training can lead to unstable training and convergence issues. A more practical approach is to first train the model solely on the recommendation loss, passing through a ``cold-start'' phase, and then introduce the unlearning terms. This approach makes it more akin to post-training unlearning.
\end{itemize}

\subsubsection{Post-training Unlearning} This category executes unlearning after the model training is completed. Post-training unlearning is generally preferred in practice because it offers greater flexibility for recommender systems to manipulate the model based on unlearning requests, without interfering with the original training process. However, post-training manipulation also presents challenges. Specifically, as training data and other training information (e.g., gradients) are typically protected or discarded after training, post-training unlearning lacks access to these resources, which limits its ability to enhance model utility.

Li et al.~\cite{li2023making} propose a bi-objective optimization framework for achieving post-training attribute unlearning. The first objective (i.e., distinguishability loss) is directly related to the primary goal of attribute unlearning. They design two types of distinguishability losses: user loss and distributional loss. The user loss manipulates the user embeddings such that users with the same attribute label are pushed apart, while users with different attribute labels are pulled together. The distributional loss treats users with the same attribute label as a group and minimizes the distance between their distributions.
The second objective (i.e., regularization loss) supports the secondary goal of attribute unlearning. Since the training data is not available during unlearning, they introduce a regularization term between the original and unlearned models to help preserve model utility.
The two objectives are combined through a linear combination, with a balancing coefficient to control the trade-off between them.

Their subsequent work further refines both objectives~\cite{chen2024post}. For the distributional loss, they introduce an anchor distribution during computation, which reduces computational complexity from exponential to linear growth with the number of classes.
For the regularization loss, they address the parameter-performance misalignment issue~\cite{benjamin2018measuring} by introducing a functional-space regularization term that better preserves original model performance without requiring training data.

Although post-training unlearning offers flexibility for unlearning execution, there is also a growing need for the unlearning process to be reversible. This is crucial, particularly when user preferences change over time or when new regulations or legal requirements emerge.
Feng et al.~\cite{feng2025plug} propose a pluggable erasure module to achieve attribute unlearning.
Specifically, they optimize through rate distortion to develop a pluggable erasure module capable of efficiently performing reversible unlearning operations, alongside attribute unlearning.
\lyy{Yu et al.~\cite{yu2025lego} propose a pluggable framework that enables dynamic unlearning of multiple attributes through a two-stage process, i.e., initial embedding calibration followed by attribute combination.}
A pluggable design is highly favored in industrial applications because it allows for more flexibility and modularity. 

Although post-training attribute unlearning methods can effectively perform unlearning without access to training data, there are two main drawbacks.
\begin{itemize}
    \item Current post-training attribute unlearning methods can only perform unlearning for all users at once, rather than on a user-specific basis. In real-world applications, recommender systems frequently need to accommodate individual user requests. 
    In contrast, in-training unlearning methods can target specific users for attribute unlearning, allowing for more fine-grained control over which users' data is affected.
    \item Training data is fundamental to building a model with stratified recommendation performance. However, post-training attribute unlearning cannot directly access this data. Existing methods primarily rely on regularization techniques applied to model parameters or recommendation rank lists. While these approaches attempt to address the absence of training data, they often lead to reduced recommendation accuracy.
    These limitations underscore the need for more effective strategies that enable attribute unlearning without compromising model performance.
\end{itemize}

\textbf{Takeaways.} To enhance readability, we summarize the focused problems, the main techniques employed, and \lyy{code reproducibility (with ``Yes'' indicating the accessible source code link)} of each method discussed above in Table~\ref{tab:all}. This table provides a concise overview, allowing a quick grasp of the core issues and approaches utilized across the various methods, offering a valuable reference for understanding the landscape of recommendation unlearning.

\begin{table*}
  \centering
  \caption{Comparison of recommendation unlearning methods.}
  \label{tab:all}
  \begin{tabular}{clllll}
    \toprule
    \textbf{Unlearning target} & \textbf{Venue} & \textbf{Focused problem} & \textbf{Main techniques} & \textbf{Model/Scenario} & \textbf{Code}\\
    \midrule
    \multirow{13}{*}{User/Item/Sample} & CSCWD'24~\cite{li2024enhancing} & Completeness & Ensemble retraining (SISA) & Agnostic & No\\
     & WWW'22~\cite{chen2022recommendation} & Completeness & Ensemble retraining & Agnostic & \href{https://github.com/chenchongthu/Recommendation-Unlearning}{Yes} \\
     & NeurIPS'23~\cite{li2023ultrare} & Efficiency, utility & Ensemble retraining, optimal transport & Agnostic & \href{https://github.com/ZhangYizhao/UltraRE}{Yes}\\
     & CIKM'23~\cite{zhang2023closed} & Efficiency & Influence function & Agnostic & No\\
     & ESWA'23~\cite{li2023selective} & Efficiency, utility & Influence function & Agnostic & No\\
     & TORS'24~\cite{zhang2023recommendation} & Efficiency & Influence function, pruning & Agnostic & \href{https://github.com/baiyimeng/IFRU}{Yes}\\
     & AAAI'24~\cite{wang2024would} & Utility & Influence function & Agnostic & \href{https://paitesanshi.github.io/IFRQE}{Yes}\\
     & KBS'24~\cite{li2024making} & Utility & Ensemble retraining, curriculum learning & Agnostic & No\\
     & AAAI'24~\cite{you2024rrl} & Utility & Loss flipping, Fisher information matrix & Agnostic & No\\
     & TOIS'24~\cite{dangefficient} & Utility & Loss flipping, multi-objective optimization & Agnostic & No\\
     & Arxiv'22~\cite{liu2022forgetting} & Efficiency & 2nd-order fine-tuning & Agnostic & No\\
     & Arxiv'23~\cite{liu2023recommendation} & Efficiency & Matrix correction & Agnostic & No\\
     & OpenReview'24~\cite{chaturvedi2024unlearning} & Efficiency & Convolutional fusion function & Agnostic & No\\
     & CIKM'24~\cite{hao2024general} & Utility & Fine-tuning, bi-objective optimization & Graph & \href{https://github.com/YongjingHao/GSGCF-RU}{Yes}\\
     & Arxiv'23~\cite{xu2023netflix} & Completeness, efficiency & Alternating least squares, fine-tuning & Bi-linear & No\\
     & Arxiv'24~\cite{sinha2024multi} & Utility, multi-modal & Loss flipping & Multi-modal & \href{https://github.com/MachineUnlearn/MMRecUN}{Yes}\\
     & Arxiv'24~\cite{hu2024exact} & Efficiency, completeness & LoRA, ensemble retraining & LLM & No\\
     & FCS'25~\cite{wang2025towards} & Efficiency, utility & LoRA, teacher-student framework & LLM & \href{https://github.com/justarter/E2URec}{Yes}\\
    \midrule
    \multirow{4}{*}{User} & WSDM'23~\cite{yuan2023federated} & Federated learning & Update calibration & Federated & No\\
     & SIGIR'23~\cite{schelter2023forget} & Efficiency & Fast indexing & KNN-based & \href{https://github.com/amsterdata/caboose}{Yes}\\
    & AJCAI'23~\cite{ye2023sequence} & Utility & Noise injection & Sequential & No \\
    & BigData'23~\cite{alshehri2023forgetting} & Utility, efficiency & Label flipping & Agnostic & No\\
    & VLDB'24~\cite{schelter2024snarcase} & Efficiency & Incremental view maintenance & Sequential & \href{https://github.com/deemdata/snapcase}{Yes}\\
    \midrule
    \multirow{2}{*}{Item} & WSDM'24~\cite{xin2024effectiveness} & Completeness & Ensemble retraining & Session-based & \href{https://github.com/shirryliu/SRU-code}{Yes}\\
     & DATA'24~\cite{wang2024forgetting} & Efficiency, utility & Rule replacement & Knowledge graph & No\\
    \midrule
    \multirow{6}{*}{Attribute} & SIGIR'22~\cite{ganhor2022unlearning} & In-training & Adversarial training & VAE & \href{https://github.com/CPJKU/adv-multvae}{Yes}\\
     & BFIR'24~\cite{escobedo2024simultaneous} & In-training, multi-attribute & Adversarial training & VAE & \href{https://github.com/hcai-mms/advx-multvae}{Yes}\\
     & ECML PKDD'24~\cite{escobedo2024making} & In-training & Item obfuscation & Agnostic & No\\
     & ACM MM'23~\cite{li2023making} & Post-training & Bi-objective optimization & Agnostic & \href{https://github.com/oktton/Attribute-wise-Unlearning}{Yes}\\
     & TOIS'24~\cite{chen2024post} & Post-training, multi-class & Bi-objective optimization & Agnostic & No\\
     & WWW'25~\cite{feng2025plug} & Post-training, pluggable & Rate distortion & Agnostic & No\\
     & ACM MM'25~\cite{yu2025lego} & Post-training, multi-attribute & Calibration and combination & Agnostic & \href{https://github.com/anonymifish/lego-rec-multiple-attribute-unlearning}{Yes}\\
    \bottomrule
  \end{tabular}
\end{table*}

\section{Evaluation}\label{sec:eva}

Based on the design principles outlined above, the evaluation of recommendation unlearning methods primarily focuses on three key aspects: unlearning completeness, unlearning efficiency, and model utility. 
Each of these aspects plays a crucial role in evaluating the effectiveness and practicality of unlearning techniques in real-world recommender systems.
In this section, we provide a comprehensive overview of the evaluation process, detailing the commonly used datasets 
and evaluation metrics employed to assess these three aspects. 

\subsection{Datasets}
Recommendation unlearning methods use the same datasets as other recommendation tasks. We list the widely used datasets and summarize the statistics in Table~\ref{tab:dataset}.

\begin{itemize}
    \item MovieLens\footnote{https://grouplens.org/datasets/movielens}: The MovieLens dataset is widely recognized as one of the most extensively used resources for recommender system research. It contains user ratings for movies and comes in multiple versions. The suffix indicates the number of interaction records; for example, MoiveLens-100K contains approximately 100,000 records.
    \item Yelp\footnote{https://www.yelp.com/dataset}: The Yelp dataset was originally compiled for the Yelp Dataset Challenge and contains users' reviews of restaurants~\cite{asghar2016yelp}. The company Yelp\footnote{https://www.yelp.com} is a platform that publishes crowd-sourced reviews of restaurants.
    which is a chance for students to conduct research or analysis on Yelp's data and share their discoveries. 
    \item Gowalla\footnote{http://snap.stanford.edu/data/loc-gowalla.html}: A location-based social networking dataset. An interaction is considered to take place whenever a user checks in at a specific location.
    \item Amazon\footnote{http://jmcauley.ucsd.edu/data/amazon}: The Amazon dataset contains several sub-datasets according to the categories of Amazon products. ADM, ELE, VG denote the sub-dataset containing reviews of digital music, electronics, and video game respectively.
    \item BookCrossing\footnote{http://www.informatik.uni-freiburg.de/\~{}cziegler/BX}: The BookCrossing dataset was collected from the Book-Crossing community and contains book ratings.
    \item Dininetica\footnote{https://darel13712.github.io/rs\_datasets/Datasets/diginetica}: The Dininetica dataset is for session-based recommendations in an e-commerce website.
    \item Steam\footnote{https://www.kaggle.com/datasets/tamber/steam-video-games/data}: The Steam dataset was collected from Steam, the world's most popular PC gaming hub, and contains transaction data for video games.
    \item MIND~\cite{wu2020mind}: The MIND dataset contains user click logs from Microsoft News, along with associated textual news information.
\end{itemize}

For attribute unlearning, the datasets must also include user attribute information in addition to the interaction records used for training. We list the representative datasets as follows:
\begin{itemize}
    \item MovieLens: Includes various attributes such as gender, age, and occupation.
    \item LFM-2B\footnote{https://www.cp.jku.at/datasets/LFM-2b}: The LFM-2B dataset contains over 2 billion listening events, designed for music retrieval and recommendation tasks~\cite{melchiorre2021investigating, schedl2022lfm}. It also includes user attributes such as gender, age, and country.
    \item KuaiSAR\footnote{https://kuaisar.github.io}: KuaiSAR is a comprehensive dataset for recommendation search, derived from user behavior logs collected from the short-video mobile application Kuaishou\footnote{https://www.kuaishou.com}. It contains various anonymous user attributes.
\end{itemize}

\begin{table}
\centering
\caption{Statistics of widely used datasets for recommendation unlearning. ``Both'' refers to the dataset being widely used for both input unlearning and attribute unlearning.}
\label{tab:dataset}
\begin{tabular}{lrrrl}  
\toprule
\textbf{Dataset} & \textbf{User \#}   & \textbf{Item \#}   & \textbf{Rating \#} & \textbf{Target}\\
\midrule
MovieLens-100K    & 943     & 1,349     & 99,287 & \multirow{3}{*}{Both}\\
MovieLens-1M    & 6,040     & 3,706     & 1,000,209 & \\
MovieLens-10M    & 71,567     & 10,681     & 10,000,054 & \\
\midrule
Yelp    & 31,668     & 38,048     & 1,561,406 & \multirow{8}{*}{Input}\\
Gowalla & 5,992 & 5,639 & 281,412 & \\
Amazon-ADM     & 478,235   & 266,414   & 836,006   & \\
Amazon-ELE     & 4,201,696 & 476,001 & 7,824,481 & \\
Amazon-VG & 28,000,000 & 1,372,000 & 46,000,000 & \\
BookCrossing & 105,283 & 340,556 & 1,149,780 & \\
Dininetica & 232,816 & 184,047 & 1,235,380 & \\
Steam & 12,393 & 5,155 & 200,000 & \\
MIND & 50,000 & 51,282 & 70,675 & \\
\midrule
LFM-2B     & 19,972   & 99,639   & 2,829,503   & \multirow{2}{*}{Attribute}\\
KuaiSAR & 21,852 & 140,367 & 2,166,893 & \\
\bottomrule
\end{tabular}
\end{table}

\subsection{Metrics}\label{sec:metric}

\subsubsection{Unlearning performance} Following the three key design principles, i.e., unlearning completeness, unlearning efficiency, and model utility, we review the commonly used evaluation metrics for each aspect. 

\paragraph{Unlearning Completeness} As mentioned in Section~\ref{sec:prin}, there are four perspectives of completeness. We introduce the metrics from each perspective as follows:
\begin{itemize}
    \item Algorithmic perspective: From this perspective, complete unlearning is defined solely as retraining the model from scratch, making the evaluation of algorithmic completeness largely theoretical. 
    \lyy{As mentioned in Section~\ref{sec:prin}, this definition can only be self-evaluated by providing training checkpoints.}
    \lyy{Thus,} exact unlearning \lyy{(i.e., retraining-based unlearning methods) naturally achieves algorithmic} completeness, while other methods do not.
    \item Parametric perspective: Although there is no literature that directly evaluates the parametric similarity between the unlearned and retrained models \lyy{in recommendation unlearning}, some research leverages Membership Inference Attacks (MIA) to assess completeness.
    In \lyy{the context of recommendation unlearning}, the input to the attacker is typically embeddings (i.e., parameters) from the recommendation model. Thus, we consider MIA evaluation as a measure of parametric completeness. 
    Specifically, MIA aims to infer whether a data record was part of the model's training set. If MIA determines that the unlearned data is no longer in the training set, unlearning is considered complete. 
    Therefore, MIA evaluation uses the attacker's performance as a metric, with typical evaluation metrics including accuracy, recall, precision, Area Under the Curve (AUC), and F1 score.
    \item Functional perspective: This perspective requires the model to perform similarly to the retrained model. Therefore, the evaluation of completeness from a functional perspective is based on recommendation performance. Representative \lyy{recommendation} metrics commonly used include Normalized Discounted Cumulative Gain (NDCG), Hit Ratio (HR)~\cite{he2015trirank}, and loss values such as Root Mean Square Error (RMSE). The evaluation is typically twofold: one on the forget set and one on the retain set. Specifically, a complete unlearning from the functional perspective should result in low performance on the \lyy{forget} set, while maintaining high performance on the \lyy{retain} set.
    \item Attack perspective: Backdoor attacks are a widely used method for evaluating completeness in traditional machine unlearning~\cite{liu2022backdoor, liu2024backdoor}. In the context of recommendation tasks, where their data consists of user-item interactions (e.g., explicit feedback with rating values or implicit feedback with binary indicators), injecting a backdoor at the sample level is challenging. 
    To address this, some research conducts poisoning attacks by flipping labels to create negative \lyy{samples}. If unlearning is completely achieved, the negative impact of \lyy{these} flipped \lyy{samples} is removed, leading to an improvement in recommendation performance, which aligns with the functional perspective of evaluation. 
    In the case of attribute unlearning, the primary goal is to make the attribute distinguishable to attackers. Therefore, completeness evaluation of attribute unlearning is typically conducted from an attack perspective, using classification metrics mentioned above, e.g., AUC and F1 score.
\end{itemize}

\paragraph{Model Utility} Model utility refers to the overall effectiveness and performance of a recommender system, particularly its ability to provide relevant recommendations after unlearning. It is typically evaluated on the testing set, which consists of unseen data that the model has not been exposed to during training and unlearning. This provides a measure of the model's ability to generalize and maintain high performance on new data.
Common evaluation metrics used to assess utility include recommendation metrics such as NDCG and HR.

\paragraph{Unlearning Efficiency} Efficiency is a crucial requirement of unlearning, particularly for large-scale models used in industry and in LLM-based recommender systems. It is essential that the unlearning process is both time-efficient and space-efficient to ensure that models remain scalable and manageable.
Both the running time and the storage cost of unlearning can be measured. Existing research primarily focuses on time efficiency, as the rapid evolution of recommender systems demands fast updates and minimal downtime.

\subsubsection{Other Deeper Impact} In addition to the common unlearning evaluation metrics mentioned above, it is important to consider the deeper, often overlooked, impacts of unlearning on the broader performance of recommender systems, particularly regarding fairness. Fairness in recommender systems is a critical issue, as it can influence how equitably the system serves different groups of users, especially when specific user data needs to be unlearned.

Li et al.~\cite{CURE4Rec2024} highlight an important concern: exact unlearning, when applied to recommender systems, can inadvertently exacerbate fairness issues. Exact unlearning typically relies on an ensemble retraining framework, where similar users (or items) are grouped together within the same data subset for retraining. 
This process leads to a phenomenon where advanced users, i.e., those with richer interaction histories, tend to be clustered together, while disadvantaged users, who may have less frequent interactions, are also grouped separately. As a result, exact unlearning can unintentionally increase the disparity between these two groups, widening the gap in fairness. The outcome is that the system may disproportionately benefit advanced users, while disadvantaging those with fewer data records or lower engagement.

In contrast, approximate unlearning does not face this issue, because it does not involve clustering of users into subsets. This is especially important in contexts where fairness is a primary concern, such as in personalized recommendations for underrepresented or marginalized user groups.

Thus, while exact unlearning offers complete data removal, it can have unintended negative consequences for fairness. Approximate unlearning, on the other hand, provides a potential solution to this issue by ensuring that fairness is not compromised during unlearning. As such, future research should explore how different unlearning approaches can balance completeness and fairness, ensuring that recommender systems serve all users equitably while respecting privacy concerns.

\section{Discussion and Future Directions}\label{sec:cha}
Based on the presented taxonomy, we provide a comprehensive summary of the similarities and differences among various recommendation unlearning methods (i.e., exact unlearning vs. approximate unlearning), highlighting the key trends and developments in existing approaches.
This summary inspires further research to explore two key questions in recommendation unlearning and, more broadly, in machine unlearning: what to unlearn? and how to unlearn?
We also compare recommendation unlearning methods with general machine unlearning techniques, examining the similarities in the design principles as well as the distinct challenges posed by the recommendation domain.
Finally, we identify several open research questions in the field, \lyy{summarizing them into four aspects, i.e., other recommendation scenarios, unlearning audition, interpretability of unlearning, and attack and defense.}

\subsection{Summary and Trends}

We summarize the characteristics and trends of existing research on recommendation unlearning and compare it with traditional machine unlearning. Our summary focuses on three key aspects: unlearning audits, unlearning methods, and emerging research tasks.

\subsubsection{Unlearning Audition (what to unlearn)}
Apart from the algorithmic auditing of unlearning completeness, most existing evaluations of unlearning primarily focus on comparing model performance rather than directly assessing the changes in model parameters. 
This shift in focus from a detailed analysis of parameter updates to a broader performance-oriented evaluation is largely driven by practical considerations in real-world scenarios.

As introduced above, in the context of recommendation unlearning, evaluations that examine unlearning from parametric and attacker perspectives are often ultimately transformed into a functional perspective. 
This transformation allows researchers and practitioners to evaluate unlearning methods in a way that is more aligned with the practical goals of maintaining model performance and data privacy while avoiding overly technical analyses that may not provide actionable insights.

Another key reason for this shift is the large parameter sizes of modern recommendation models, particularly the user and item embeddings. These embeddings typically involve millions (or even billions) of parameters, making it exceedingly difficult to directly compare the changes in parameters after unlearning. The sheer scale of these models means that tracking and analyzing individual parameter adjustments not only incurs significant computational overhead but also becomes intractable.

Moreover, direct parameter comparison might not always provide meaningful insights into the completeness of unlearning. For example, even if the parameters change significantly, it may not necessarily reflect a corresponding improvement in privacy. Instead, focusing on the functional performance offers a more holistic view of how unlearning techniques impact the system.

\textbf{In summary}, while parameter-based evaluations provide a granular view of the unlearning process, their practical challenges, especially for large-scale recommender systems, have led to a greater emphasis on functional evaluations that focus on model performance. As recommendation models continue to grow in complexity, the need for more efficient, high-level evaluations will only increase.

\subsubsection{Unlearning Methods (how to unlearn)}
In the context of unlearning in recommender systems, both exact and approximate unlearning approaches have been explored, with approximate methods gaining increasing popularity. This trend aligns with the ongoing developments in traditional machine unlearning research. 
The growing preference for approximate unlearning can be attributed to several factors, primarily the trade-off between theoretical guarantees and empirical performance. While exact unlearning provides a strong theoretical foundation for ensuring complete unlearning of forget set, its empirical performance often lags behind that of approximate methods~\cite{CURE4Rec2024}. This is particularly evident in real-world applications, where users typically prioritize practical performance (e.g., model utility and efficiency) over theoretical guarantees.

Exact unlearning requires retraining models from scratch, which imposes significant computational and time costs. This strict requirement severely constrains the applicability of exact unlearning in large-scale systems, where the need for rapid, on-the-fly updates is paramount. 

In contrast, approximate unlearning allows for more flexibility in handling unlearning.
The broader scope of techniques available for approximate unlearning inspires new insights and research directions. These methods allow researchers to explore innovative ways in the future. 
Additionally, the issue of unfairness leads to an increased focus on developing methods that not only meet the privacy requirements but also ensure that the system continues to perform well across all user groups, regardless of their data history.

\textbf{In summary}, while exact unlearning provides robust theoretical guarantees, the empirical challenges it presents in terms of efficiency and utility make approximate unlearning a more attractive choice for real-world recommender systems. As research in this area continues to evolve, the exploration of new approximate unlearning techniques will likely remain a key area of focus, driving improvements in both system performance and data privacy.

\subsubsection{\lyy{Emerging} Research Tasks}
Recommendation unlearning has spurred fresh perspectives in the broader unlearning research field, such as the emergence of attribute unlearning and LLM-based recommendation unlearning. These innovations underscore the expanding diversity of challenges and opportunities in this area. 
On the one hand, there is an increasing body of work focusing on unlearning techniques tailored to specific types of recommendation models, addressing the unique characteristics and complexities of various algorithms, such as graph-based models and KNN-based recommendation models. 
On the other hand, there is also a growing focus on unlearning within diverse recommendation scenarios. This includes federated learning, sequential recommendation, session-based recommendations, and LLMs.

\lyy{\textbf{In summary}}, these advancements highlight the growing need for unlearning methodologies that are not only model-specific but also adaptable to a broader range of use cases. As recommender systems become more sophisticated and pervasive, investigating recommendation unlearning across a variety of settings will be essential. The ongoing evolution of unlearning techniques promises to offer more robust, scalable solutions for handling sensitive and dynamic user data, while maintaining the functionality and fairness of recommender systems in increasingly complex environments.

\subsection{Open Research Questions}

\subsubsection{Other Recommendation Scenarios}
In addition to the traditional recommendation scenarios, existing studies have begun to explore more recommendation scenarios. Below, we outline a few of these scenarios that require further attention:
\begin{itemize}
    \item Sequential unlearning. In real-world applications, unlearning requests from different users often arrive sequentially. 
    However, traditional recommendation unlearning methods process requests in batch mode, which, while effective in certain contexts, is impractical in dynamic environments where responsiveness is critical. 
    In this sequential setting, exact unlearning, which requires frequent model retraining, becomes inefficient and may essentially be reduced to a condition where they must be retrained from scratch. 
    On the other hand, approximate unlearning faces a unique challenge in sequential settings: the estimation gap becomes more pronounced over time, making it more difficult to maintain utility and completeness in the recommendations.
    Therefore, the scenario of sequential unlearning remains worthy of further investigation.
    \item User-perspective unlearning. Most existing unlearning research adopts a system-centric perspective, where the recommender system itself is responsible for performing the unlearning operation. However, there is an emerging need to explore user-perspective unlearning, where users can proactively manage their data by requesting the deletion or modification of their preferences, without necessarily being dependent on the system to do so. 
    \lyy{Several studies in the human-computer interaction community have explored similar concepts of user-perspective actions~\cite{liu2025filtering, shu2024rah}.}
    This would enable users to erase their personal data or adjust their preferences autonomously, even in systems where the platform does not explicitly support such features. User-perspective unlearning has significant implications for user privacy and control, as it allows users to take ownership of their data and make unlearning requests directly, without waiting for the system to initiate or support the process. This scenario raises new challenges around how to empower users to manage their own data effectively, especially in platforms that are not inherently designed for such dynamic interactions.
\end{itemize}

\subsubsection{Unlearning Audition}
One of the central challenges in unlearning research is developing a standard, transferable method for evaluating completeness. This is particularly important in the context of sample-wise unlearning (i.e., unlearning a single piece of user-item interaction record). \lyy{However, the memory or contribution of a single data point can be difficult to assess}
\lyy{On the one hand}, removing one or a few data points might not substantially affect the model’s accuracy, loss, or parameters due to the \lyy{re-remember phenomenon illustrated in Figure~\ref{fig:example}}.
\lyy{On the other hand, a small number of vital interactions can drastically impact particular embedding and recommendations due to the}
influence of other data points. Thus, developing metrics to accurately quantify the influence of sample-wise unlearning is critical.
Alternatively, it may be impossible to assess sample-wise influence, with sample-wise unlearning existing only at an algorithmic level. This question is also present in traditional machine unlearning research.

\subsubsection{Interpretability of Unlearning} The interpretability of unlearning is a relatively underexplored area in the literature.
\begin{itemize}
    \item Understanding the effect of unlearning techniques on both completeness and utility remains an open question. For example, techniques such as SISA aim to improve unlearning efficiency, but their impact on utility needs further exploration. A key research direction involves interpreting how unlearning influences model behavior and ensuring that it can be done in a way that is both transparent and beneficial for end-users.
    \item What does a recommendation model learn, and what should be unlearned? Given the stochastic nature and randomness inherent in deep learning, this remains an open question. 
    Prior studies provide definitions from an algorithmic perspective~\cite{jia2021proof, thudi2022necessity, weng2024proof}, but a more comprehensive definition that incorporates both the parametric and functional levels is desired. Such an approach would contribute significantly to the interpretability of both the learning and unlearning processes.
\end{itemize}

\subsubsection{Attack and Defense}
As unlearning techniques continue to evolve, so too do the potential vulnerabilities they may introduce~\cite{leysen2023exploring}. While unlearning is designed to eliminate the influence of specific data points, it is not immune to adversarial threats. Robust unlearning must, therefore, incorporate strategies to safeguard the model from attacks that aim to exploit weaknesses in the unlearning process itself.

Moreover, unlearning can inadvertently create a false sense of security. In fact, unlearning the data that are perceived to be harmful or unnecessary may exaggerate the model's vulnerability to future attacks.
For example, the memorization of deep learning models exhibits an \textit{onion effect}, where removing ``outliers'' from the training data may inadvertently increase the vulnerability of the ``inner data'' to attacks~\cite{carlini2022privacy}.
Furthermore, attackers can exploit the differences between the model parameters before and after unlearning to design more sophisticated and effective attacks~\cite{chen2021machine}.

Thus, developing techniques that ensure unlearning is both secure and resilient against adversarial attacks represents a significant challenge for future research. These techniques must not only focus on the technical integrity of the unlearning process but also account for the broader security implications.

\section{Conclusion}
In this paper, we provide a comprehensive review of recommendation unlearning. We first introduce the foundational concepts, followed by a unified taxonomy. This taxonomy highlights the diverse approaches proposed to address different unlearning targets in recommender systems (i.e., input unlearning and attribute unlearning), offering a clearer understanding of the underlying strategies (e.g., exact unlearning, reverse unlearning, and active unlearning) with their respective strengths and limitations.
We also present evaluation metrics based on three key design principles: unlearning completeness, unlearning efficiency, and model utility. In this context, we introduce widely used datasets and models to facilitate research in the field.
Finally, we outline several open research questions not only related to unlearning in recommender systems but also applicable to broader scenarios. These questions call for further exploration of areas such as unlearning auditing, interpretability, attack, and defense, which pave the way for future advancements that could enhance both the practical utility and ethical considerations of recommendation unlearning and, more broadly, machine unlearning.

\section*{Acknowledgment}
This work was supported in part by the National Natural Science Foundation of China (No.~62402148).


\bibliographystyle{named}
\bibliography{sample-base}

@String{Computing = "Computing" }

@String{Computer = "{IEEE} Computer" }

@String{Springer = "Springer-Verlag" }

@inproceedings{he2016deep,
  title={Deep residual learning for image recognition},
  author={He, Kaiming and Zhang, Xiangyu and Ren, Shaoqing and Sun, Jian},
  booktitle={Proceedings of the IEEE conference on computer vision and pattern recognition (CVPR)},
  pages={770--778},
  year={2016}
}

@article{lecun2015deep,
  title={Deep learning},
  author={LeCun, Yann and Bengio, Yoshua and Hinton, Geoffrey},
  journal={nature},
  volume={521},
  number={7553},
  pages={436--444},
  year={2015},
  publisher={Nature Publishing Group UK London}
}

@article{openai2023gpt,
  author    = {OpenAI},
  title     = {{GPT-4} Technical Report},
  journal   = {CoRR},
  volume    = {abs/2303.08774},
  year      = {2023},
  url       = {https://doi.org/10.48550/arXiv.2303.08774},
  doi       = {10.48550/arXiv.2303.08774},
  eprinttype = {arXiv},
  eprint    = {2303.08774},
  bibsource = {dblp computer science bibliography, https://dblp.org}
}

@misc{euro2018gdpr,
    author = {European Union},
    title = {General Data Protection Regulation},
    year = {2018},
    url = {https://gdpr-info.eu/}
}

@misc{cal2023del,
    author = {California Legislative Information},
    title = {California Senate Bill 362},
    year = {2023},
    url = {https://leginfo.legislature.ca.gov/faces/billNavClient.xhtml?bill_id=202120220SB362}
}

@misc{cal2018ccpa,
    author = {State of California Department of Justice},
    title = {California Consumer Privacy Act (CCPA)},
    year = {2018},
    url = {https://oag.ca.gov/privacy/ccpa}
}

@inproceedings{cao2015towards,
  title={Towards Making Systems Forget with Machine Unlearning},
  author={Cao, Yinzhi and Yang, Junfeng},
  booktitle={Proceedings in the 36th IEEE Symposium on Security and Privacy},
  pages={463--480},
  year={2015}
}

@inproceedings{steck2013evaluation,
  title={Evaluation of recommendations: rating-prediction and ranking},
  author={Steck, Harald},
  booktitle={Proceedings of the 7th ACM conference on Recommender systems},
  pages={213--220},
  year={2013}
}

@article{koren2021advances,
  title={Advances in collaborative filtering},
  author={Koren, Yehuda and Rendle, Steffen and Bell, Robert},
  journal={Recommender systems handbook},
  pages={91--142},
  year={2021},
  publisher={Springer}
}

@inproceedings{ma2008sorec,
  title={Sorec: social recommendation using probabilistic matrix factorization},
  author={Ma, Hao and Yang, Haixuan and Lyu, Michael R and King, Irwin},
  booktitle={Proceedings of the 17th ACM conference on Information and knowledge management},
  pages={931--940},
  year={2008}
}

@inproceedings{hu2008collaborative,
  title={Collaborative filtering for implicit feedback datasets},
  author={Hu, Yifan and Koren, Yehuda and Volinsky, Chris},
  booktitle={2008 Eighth IEEE international conference on data mining},
  pages={263--272},
  year={2008},
  organization={Ieee}
}

@inproceedings{xue2017deep,
  title={Deep Matrix Factorization Models for Recommender Systems.},
  author={Xue, Hong-Jian and Dai, Xinyu and Zhang, Jianbing and Huang, Shujian and Chen, Jiajun},
  booktitle={Proceedings of the 26th International Joint Conference on Artificial Intelligence (IJCAI)},
  volume={17},
  pages={3203--3209},
  year={2017},
  organization={Melbourne, Australia}
}

@inproceedings{peng2022svd,
  title={SVD-GCN: A simplified graph convolution paradigm for recommendation},
  author={Peng, Shaowen and Sugiyama, Kazunari and Mine, Tsunenori},
  booktitle={Proceedings of the 31st ACM international conference on information \& knowledge management},
  pages={1625--1634},
  year={2022}
}

@inproceedings{zheng2021dgcn,
  title={DGCN: Diversified recommendation with graph convolutional networks},
  author={Zheng, Yu and Gao, Chen and Chen, Liang and Jin, Depeng and Li, Yong},
  booktitle={Proceedings of the Web Conference 2021},
  pages={401--412},
  year={2021}
}

@inproceedings{zhao2021feasibility,
  title={On the (in) feasibility of attribute inference attacks on machine learning models},
  author={Zhao, Benjamin Zi Hao and Agrawal, Aviral and Coburn, Catisha and Asghar, Hassan Jameel and Bhaskar, Raghav and Kaafar, Mohamed Ali and Webb, Darren and Dickinson, Peter},
  booktitle={2021 IEEE European Symposium on Security and Privacy (EuroS\&P)},
  pages={232--251},
  year={2021},
  organization={IEEE}
}

@article{gong2018attribute,
  title={Attribute inference attacks in online social networks},
  author={Gong, Neil Zhenqiang and Liu, Bin},
  journal={ACM Transactions on Privacy and Security (TOPS)},
  volume={21},
  number={1},
  pages={1--30},
  year={2018},
  publisher={ACM New York, NY, USA}
}

@inproceedings{jayaraman2022attribute,
  title={Are attribute inference attacks just imputation?},
  author={Jayaraman, Bargav and Evans, David},
  booktitle={Proceedings of the 2022 ACM SIGSAC Conference on Computer and Communications Security},
  pages={1569--1582},
  year={2022}
}

@article{sachdeva2024machine,
  title={Machine Unlearning for Recommendation Systems: An Insight},
  author={Sachdeva, Bhavika and Rathee, Harshita and Sharma, Arun and Wydma{\'n}ski, Witold and others},
  journal={arXiv preprint arXiv:2401.10942},
  year={2024}
}

@inproceedings{chen2022recommendation,
  title={Recommendation unlearning},
  author={Chen, Chong and Sun, Fei and Zhang, Min and Ding, Bolin},
  booktitle={Proceedings of the ACM Web Conference 2022},
  pages={2768--2777},
  year={2022}
}

@article{li2023selective,
title = {Selective and collaborative influence function for efficient recommendation unlearning},
journal = {Expert Systems with Applications},
volume = {234},
pages = {121025},
year = {2023},
issn = {0957-4174},
author = {Li, Yuyuan and Chen, Chaochao and Zheng, Xiaolin and Zhang, Yizhao and Gong, Biao and Wang, Jun and Chen, Linxun}
}

@article{li2024making,
  title={Making recommender systems forget: Learning and unlearning for erasable recommendation},
  author={Li, Yuyuan and Chen, Chaochao and Zheng, Xiaolin and Liu, Junlin and Wang, Jun},
  journal={Knowledge-Based Systems},
  volume={283},
  pages={111124},
  year={2024},
  publisher={Elsevier}
}

@article{li2023ultrare,
  title={Ultrare: Enhancing receraser for recommendation unlearning via error decomposition},
  author={Li, Yuyuan and Chen, Chaochao and Zhang, Yizhao and Liu, Weiming and Lyu, Lingjuan and Zheng, Xiaolin and Meng, Dan and Wang, Jun},
  journal={Advances in Neural Information Processing Systems},
  volume={36},
  year={2023}
}

@article{liu2022forgetting,
  title={Forgetting fast in recommender systems},
  author={Liu, Wenyan and Wan, Juncheng and Wang, Xiaoling and Zhang, Weinan and Zhang, Dell and Li, Hang},
  journal={arXiv preprint arXiv:2208.06875},
  year={2022}
}

@inproceedings{yuan2023federated,
  title={Federated unlearning for on-device recommendation},
  author={Yuan, Wei and Yin, Hongzhi and Wu, Fangzhao and Zhang, Shijie and He, Tieke and Wang, Hao},
  booktitle={Proceedings of the Sixteenth ACM International Conference on Web Search and Data Mining},
  pages={393--401},
  year={2023}
}

@article{xu2023netflix,
  title={Netflix and forget: Efficient and exact machine unlearning from bi-linear recommendations},
  author={Xu, Mimee and Sun, Jiankai and Yang, Xin and Yao, Kevin and Wang, Chong},
  journal={arXiv preprint arXiv:2302.06676},
  year={2023}
}

@article{liu2023recommendation,
  title={Recommendation Unlearning via Matrix Correction},
  author={Liu, Jiahao and Li, Dongsheng and Gu, Hansu and Lu, Tun and Wu, Jiongran and Zhang, Peng and Shang, Li and Gu, Ning},
  journal={arXiv preprint arXiv:2307.15960},
  year={2023}
}

@article{zhang2023recommendation,
  title={Recommendation unlearning via influence function},
  author={Zhang, Yang and Hu, Zhiyu and Bai, Yimeng and Wu, Jiancan and Wang, Qifan and Feng, Fuli},
  journal={ACM Transactions on Recommender Systems},
  year={2024},
  publisher={ACM New York, NY}
}

@inproceedings{schelter2023forget,
  title={Forget me now: Fast and exact unlearning in neighborhood-based recommendation},
  author={Schelter, Sebastian and Ariannezhad, Mozhdeh and de Rijke, Maarten},
  booktitle={Proceedings of the 46th International ACM SIGIR Conference on Research and Development in Information Retrieval},
  pages={2011--2015},
  year={2023}
}

@inproceedings{li2023making,
  title={Making users indistinguishable: Attribute-wise unlearning in recommender systems},
  author={Li, Yuyuan and Chen, Chaochao and Zheng, Xiaolin and Zhang, Yizhao and Han, Zhongxuan and Meng, Dan and Wang, Jun},
  booktitle={Proceedings of the 31st ACM International Conference on Multimedia},
  pages={984--994},
  year={2023}
}

@inproceedings{ganhor2022unlearning,
  title={Unlearning protected user attributes in recommendations with adversarial training},
  author={Ganh{\"o}r, Christian and Penz, David and Rekabsaz, Navid and Lesota, Oleg and Schedl, Markus},
  booktitle={Proceedings of the 45th International ACM SIGIR Conference on Research and Development in Information Retrieval},
  pages={2142--2147},
  year={2022}
}

@inproceedings{ye2023sequence,
  title={Sequence Unlearning for Sequential Recommender Systems},
  author={Ye, Shanshan and Lu, Jie},
  booktitle={Australasian Joint Conference on Artificial Intelligence},
  pages={403--415},
  year={2023},
  organization={Springer}
}

@inproceedings{xin2024effectiveness,
  title={On the effectiveness of unlearning in session-based recommendation},
  author={Xin, Xin and Yang, Liu and Zhao, Ziqi and Ren, Pengjie and Chen, Zhumin and Ma, Jun and Ren, Zhaochun},
  booktitle={Proceedings of the 17th ACM International Conference on Web Search and Data Mining},
  pages={855--863},
  year={2024}
}

@article{vaswani2017attention,
  title={Attention is all you need},
  author={Vaswani, Ashish and Shazeer, Noam and Parmar, Niki and Uszkoreit, Jakob and Jones, Llion and Gomez, Aidan N and Kaiser, {\L}ukasz and Polosukhin, Illia},
  journal={Advances in neural information processing systems},
  volume={30},
  year={2017}
}

@inproceedings{bourtoule2021machine,
  title={Machine unlearning},
  author={Bourtoule, Lucas and Chandrasekaran, Varun and Choquette-Choo, Christopher A and Jia, Hengrui and Travers, Adelin and Zhang, Baiwu and Lie, David and Papernot, Nicolas},
  booktitle={2021 IEEE Symposium on Security and Privacy (SP)},
  pages={141--159},
  year={2021},
  organization={IEEE}
}

@article{li2016data,
  title={Data poisoning attacks on factorization-based collaborative filtering},
  author={Li, Bo and Wang, Yining and Singh, Aarti and Vorobeychik, Yevgeniy},
  journal={Advances in Neural Information Processing Systems (NeurIPS)},
  volume={29},
  year={2016}
}

@article{mehrabi2021survey,
  title={A survey on bias and fairness in machine learning},
  author={Mehrabi, Ninareh and Morstatter, Fred and Saxena, Nripsuta and Lerman, Kristina and Galstyan, Aram},
  journal={ACM Computing Surveys (CSUR)},
  volume={54},
  number={6},
  pages={1--35},
  year={2021},
  publisher={ACM New York, NY, USA}
}

@article{chen2024post,
  title={Post-training attribute unlearning in recommender systems},
  author={Chen, Chaochao and Zhang, Yizhao and Li, Yuyuan and Wang, Jun and Qi, Lianyong and Xu, Xiaolong and Zheng, Xiaolin and Yin, Jianwei},
  journal={ACM Transactions on Information Systems},
  year={2024},
  publisher={ACM New York, NY}
}

@inproceedings{you2024rrl,
  title={RRL: Recommendation Reverse Learning},
  author={You, Xiaoyu and Xu, Jianwei and Zhang, Mi and Gao, Zechen and Yang, Min},
  booktitle={Proceedings of the AAAI Conference on Artificial Intelligence},
  volume={38},
  number={8},
  pages={9296--9304},
  year={2024}
}

@inproceedings{alshehri2023forgetting,
  title={Forgetting User Preference in Recommendation Systems with Label-Flipping},
  author={Alshehri, Manal A and Zhang, Xiangliang},
  booktitle={2023 IEEE International Conference on Big Data (BigData)},
  pages={271--281},
  year={2023},
  organization={IEEE}
}

@inproceedings{escobedo2024making,
  title={Making Alice Appear Like Bob: A Probabilistic Preference Obfuscation Method For Implicit Feedback Recommendation Models},
  author={Escobedo, Gustavo and Moscati, Marta and Muellner, Peter and Kopeinik, Simone and Kowald, Dominik and Lex, Elisabeth and Schedl, Markus},
  booktitle={Joint European Conference on Machine Learning and Knowledge Discovery in Databases},
  pages={349--365},
  year={2024},
  organization={Springer}
}

@article{sinha2024multi,
  title={Multi-Modal Recommendation Unlearning},
  author={Sinha, Yash and Mandal, Murari and Kankanhalli, Mohan},
  journal={arXiv preprint arXiv:2405.15328},
  year={2024}
}

@inproceedings{leysen2023exploring,
  title={Exploring unlearning methods to ensure the privacy, security, and usability of recommender systems},
  author={Leysen, Jens},
  booktitle={Proceedings of the 17th ACM Conference on Recommender Systems},
  pages={1300--1304},
  year={2023}
}

@inproceedings{wang2024forgetting,
  title={Forgetting in Knowledge Graph Based Recommender Systems},
  author={Wang, Xu and Brewster, Christopher},
  booktitle={Proceedings of the 13th International Conference on Data Science, Technology and Applications},
  pages={309--317},
  year={2024}
}

@inproceedings{wang2024would,
  title={Would You Like Your Data to Be Trained? A User Controllable Recommendation Framework},
  author={Wang, Lei and Chen, Xu and Dong, Zhenhua and Dai, Quanyu},
  booktitle={Proceedings of the AAAI Conference on Artificial Intelligence},
  volume={38},
  number={19},
  pages={21673--21680},
  year={2024}
}

@misc{chaturvedi2024unlearning,
  title={Unlearning the Unwanted Data from a Personalized Recommendation Model},
  author={Chaturvedi, Narayan and Singh, Brijraj and Pedanekar, Niranjan},
  year={2024},
  url={https://openreview.net/pdf?id=3Ok7ccvtf3}
}

@article{hu2024exact,
  title={Exact and Efficient Unlearning for Large Language Model-based Recommendation},
  author={Hu, Zhiyu and Zhang, Yang and Xiao, Minghao and Wang, Wenjie and Feng, Fuli and He, Xiangnan},
  journal={arXiv preprint arXiv:2404.10327},
  year={2024}
}

@article{ho2020denoising,
  title={Denoising diffusion probabilistic models},
  author={Ho, Jonathan and Jain, Ajay and Abbeel, Pieter},
  journal={Advances in neural information processing systems},
  volume={33},
  pages={6840--6851},
  year={2020}
}

@inproceedings{he2020lightgcn,
  title={Lightgcn: Simplifying and powering graph convolution network for recommendation},
  author={He, Xiangnan and Deng, Kuan and Wang, Xiang and Li, Yan and Zhang, Yongdong and Wang, Meng},
  booktitle={Proceedings of the 43rd International ACM SIGIR conference on research and development in Information Retrieval},
  pages={639--648},
  year={2020}
}

@inproceedings{wang2019neural,
  title={Neural graph collaborative filtering},
  author={Wang, Xiang and He, Xiangnan and Wang, Meng and Feng, Fuli and Chua, Tat-Seng},
  booktitle={Proceedings of the 42nd international ACM SIGIR conference on Research and development in Information Retrieval},
  pages={165--174},
  year={2019}
}

@inproceedings{he2017neural,
  title={Neural collaborative filtering},
  author={He, Xiangnan and Liao, Lizi and Zhang, Hanwang and Nie, Liqiang and Hu, Xia and Chua, Tat-Seng},
  booktitle={Proceedings of the 26th international conference on world wide web},
  pages={173--182},
  year={2017}
}

@article{chen2020efficient,
  title={Efficient neural matrix factorization without sampling for recommendation},
  author={Chen, Chong and Zhang, Min and Zhang, Yongfeng and Liu, Yiqun and Ma, Shaoping},
  journal={ACM Transactions on Information Systems (TOIS)},
  volume={38},
  number={2},
  pages={1--28},
  year={2020},
  publisher={ACM New York, NY, USA}
}

@article{CURE4Rec2024,
    title={CURE4Rec: A Benchmark for Recommendation Unlearning with Deeper Influence}, 
    author={Chen, Chaochao and Zhang, Jiaming and Zhang, Yizhao and Zhang, Li and Lyu, Lingjuan and Li, Yuyuan and Gong, Biao and Yan, Chenggang},
    journal={Advances in Neural Information Processing Systems},
    volume={37},
    year={2024},  
}

@inproceedings{hao2024general,
  title={A General Strategy Graph Collaborative Filtering for Recommendation Unlearning},
  author={Hao, Yongjing and Zhuang, Fuzhen and Wang, Deqing and Liu, Guanfeng and Sheng, Victor S and Zhao, Pengpeng},
  booktitle={Proceedings of the 33rd ACM International Conference on Information and Knowledge Management},
  pages={799--808},
  year={2024}
}

@inproceedings{zhang2023closed,
  title={Closed-form machine unlearning for matrix factorization},
  author={Zhang, Shuijing and Lou, Jian and Xiong, Li and Zhang, Xiaoyu and Liu, Jing},
  booktitle={Proceedings of the 32nd ACM International Conference on Information and Knowledge Management},
  pages={3278--3287},
  year={2023}
}

@inproceedings{li2024enhancing,
  title={Enhancing Privacy Protection for Online Learning Resource Recommendation with Machine Unlearning},
  author={Li, Wenqin and Zheng, Xinrong and Huang, Ruihong and Lin, Mingwei and Shen, Jun and Lin, Jiayin},
  booktitle={2024 27th International Conference on Computer Supported Cooperative Work in Design (CSCWD)},
  pages={3282--3287},
  year={2024},
  organization={IEEE}
}

@article{zheng2022matrix,
  title={A matrix factorization recommendation system-based local differential privacy for protecting users’ sensitive data},
  author={Zheng, Xiaoyao and Guan, Manping and Jia, Xianmin and Guo, Liangmin and Luo, Yonglong},
  journal={IEEE Transactions on Computational Social Systems},
  volume={10},
  number={3},
  pages={1189--1198},
  year={2022},
  publisher={IEEE}
}

@incollection{schafer2007collaborative,
  title={Collaborative filtering recommender systems},
  author={Schafer, J Ben and Frankowski, Dan and Herlocker, Jon and Sen, Shilad},
  booktitle={The adaptive web},
  pages={291--324},
  year={2007},
  publisher={Springer}
}

@article{nguyen2022survey,
  title={A survey of machine unlearning},
  author={Nguyen, Thanh Tam and Huynh, Thanh Trung and Nguyen, Phi Le and Liew, Alan Wee-Chung and Yin, Hongzhi and Nguyen, Quoc Viet Hung},
  journal={arXiv preprint arXiv:2209.02299},
  year={2022}
}

@inproceedings{biderman2023pythia,
  title={Pythia: A suite for analyzing large language models across training and scaling},
  author={Biderman, Stella and Schoelkopf, Hailey and Anthony, Quentin Gregory and Bradley, Herbie and O’Brien, Kyle and Hallahan, Eric and Khan, Mohammad Aflah and Purohit, Shivanshu and Prashanth, USVSN Sai and Raff, Edward and others},
  booktitle={International Conference on Machine Learning},
  pages={2397--2430},
  year={2023},
  organization={PMLR}
}

@article{touvron2023llama,
  title={Llama 2: Open foundation and fine-tuned chat models},
  author={Touvron, Hugo and Martin, Louis and Stone, Kevin and Albert, Peter and Almahairi, Amjad and Babaei, Yasmine and Bashlykov, Nikolay and Batra, Soumya and Bhargava, Prajjwal and Bhosale, Shruti and others},
  journal={arXiv preprint arXiv:2307.09288},
  year={2023}
}

@article{dubey2024llama,
  title={The llama 3 herd of models},
  author={Dubey, Abhimanyu and Jauhri, Abhinav and Pandey, Abhinav and Kadian, Abhishek and Al-Dahle, Ahmad and Letman, Aiesha and Mathur, Akhil and Schelten, Alan and Yang, Amy and Fan, Angela and others},
  journal={arXiv preprint arXiv:2407.21783},
  year={2024}
}

@article{dettmers2024qlora,
  title={Qlora: Efficient finetuning of quantized llms},
  author={Dettmers, Tim and Pagnoni, Artidoro and Holtzman, Ari and Zettlemoyer, Luke},
  journal={Advances in Neural Information Processing Systems},
  volume={36},
  year={2024}
}

@inproceedings{NEURIPS2020_1457c0d6,
 author = {Brown, Tom and Mann, Benjamin and Ryder, Nick and Subbiah, Melanie and Kaplan, Jared D and Dhariwal, Prafulla and Neelakantan, Arvind and Shyam, Pranav and Sastry, Girish and Askell, Amanda and Agarwal, Sandhini and Herbert-Voss, Ariel and Krueger, Gretchen and Henighan, Tom and Child, Rewon and Ramesh, Aditya and Ziegler, Daniel and Wu, Jeffrey and Winter, Clemens and Hesse, Chris and Chen, Mark and Sigler, Eric and Litwin, Mateusz and Gray, Scott and Chess, Benjamin and Clark, Jack and Berner, Christopher and McCandlish, Sam and Radford, Alec and Sutskever, Ilya and Amodei, Dario},
 booktitle = {Advances in Neural Information Processing Systems},
 pages = {1877--1901},
 publisher = {Curran Associates, Inc.},
 title = {Language Models are Few-Shot Learners},
 volume = {33},
 year = {2020}
}

@article{radford2019language,
  title={Language models are unsupervised multitask learners},
  author={Radford, Alec and Wu, Jeffrey and Child, Rewon and Luan, David and Amodei, Dario and Sutskever, Ilya and others},
  journal={OpenAI blog},
  volume={1},
  number={8},
  pages={9},
  year={2019}
}

@article{wu2023bloomberggpt,
  title={Bloomberggpt: A large language model for finance},
  author={Wu, Shijie and Irsoy, Ozan and Lu, Steven and Dabravolski, Vadim and Dredze, Mark and Gehrmann, Sebastian and Kambadur, Prabhanjan and Rosenberg, David and Mann, Gideon},
  journal={arXiv preprint arXiv:2303.17564},
  year={2023}
}

@article{jumper2021highly,
  title={Highly accurate protein structure prediction with AlphaFold},
  author={Jumper, John and Evans, Richard and Pritzel, Alexander and Green, Tim and Figurnov, Michael and Ronneberger, Olaf and Tunyasuvunakool, Kathryn and Bates, Russ and {\v{Z}}{\'\i}dek, Augustin and Potapenko, Anna and others},
  journal={nature},
  volume={596},
  number={7873},
  pages={583--589},
  year={2021},
  publisher={Nature Publishing Group}
}

@article{silver2016mastering,
  title={Mastering the game of Go with deep neural networks and tree search},
  author={Silver, David and Huang, Aja and Maddison, Chris J and Guez, Arthur and Sifre, Laurent and Van Den Driessche, George and Schrittwieser, Julian and Antonoglou, Ioannis and Panneershelvam, Veda and Lanctot, Marc and others},
  journal={nature},
  volume={529},
  number={7587},
  pages={484--489},
  year={2016},
  publisher={Nature Publishing Group}
}

@article{silver2017mastering,
  title={Mastering the game of go without human knowledge},
  author={Silver, David and Schrittwieser, Julian and Simonyan, Karen and Antonoglou, Ioannis and Huang, Aja and Guez, Arthur and Hubert, Thomas and Baker, Lucas and Lai, Matthew and Bolton, Adrian and others},
  journal={nature},
  volume={550},
  number={7676},
  pages={354--359},
  year={2017},
  publisher={Nature Publishing Group}
}

@article{abramson2024accurate,
  title={Accurate structure prediction of biomolecular interactions with AlphaFold 3},
  author={Abramson, Josh and Adler, Jonas and Dunger, Jack and Evans, Richard and Green, Tim and Pritzel, Alexander and Ronneberger, Olaf and Willmore, Lindsay and Ballard, Andrew J and Bambrick, Joshua and others},
  journal={Nature},
  pages={1--3},
  year={2024},
  publisher={Nature Publishing Group UK London}
}

@article{stahl2018ethics,
  title={Ethics and privacy in AI and big data: Implementing responsible research and innovation},
  author={Stahl, Bernd Carsten and Wright, David},
  journal={IEEE Security \& Privacy},
  volume={16},
  number={3},
  pages={26--33},
  year={2018},
  publisher={IEEE}
}

@article{eshete2021making,
  title={Making machine learning trustworthy},
  author={Eshete, Birhanu},
  journal={Science},
  volume={373},
  number={6556},
  pages={743--744},
  year={2021},
  publisher={American Association for the Advancement of Science}
}

@inproceedings{zhang2024instruction,
  title={Instruction backdoor attacks against customized $\{$LLMs$\}$},
  author={Zhang, Rui and Li, Hongwei and Wen, Rui and Jiang, Wenbo and Zhang, Yuan and Backes, Michael and Shen, Yun and Zhang, Yang},
  booktitle={33rd USENIX Security Symposium (USENIX Security 24)},
  pages={1849--1866},
  year={2024}
}

@article{wang2024machine,
  title={Machine unlearning: A comprehensive survey},
  author={Wang, Weiqi and Tian, Zhiyi and Zhang, Chenhan and Yu, Shui},
  journal={arXiv preprint arXiv:2405.07406},
  year={2024}
}

@article{liu2024threats,
  title={Threats, attacks, and defenses in machine unlearning: A survey},
  author={Liu, Ziyao and Ye, Huanyi and Chen, Chen and Zheng, Yongsen and Lam, Kwok-Yan},
  journal={arXiv preprint arXiv:2403.13682},
  year={2024}
}

@article{shaik2024exploring,
  title={Exploring the landscape of machine unlearning: A comprehensive survey and taxonomy},
  author={Shaik, Thanveer and Tao, Xiaohui and Xie, Haoran and Li, Lin and Zhu, Xiaofeng and Li, Qing},
  journal={IEEE Transactions on Neural Networks and Learning Systems},
  year={2024},
  publisher={IEEE}
}

@article{liu2024machine,
  title={Machine unlearning in generative ai: A survey},
  author={Liu, Zheyuan and Dou, Guangyao and Tan, Zhaoxuan and Tian, Yijun and Jiang, Meng},
  journal={arXiv preprint arXiv:2407.20516},
  year={2024}
}

@article{li2024machine,
  title={Machine Unlearning: Taxonomy, Metrics, Applications, Challenges, and Prospects},
  author={Li, Na and Zhou, Chunyi and Gao, Yansong and Chen, Hui and Fu, Anmin and Zhang, Zhi and Shui, Yu},
  journal={arXiv preprint arXiv:2403.08254},
  year={2024}
}

@article{liu2024survey,
  title={A survey on federated unlearning: Challenges, methods, and future directions},
  author={Liu, Ziyao and Jiang, Yu and Shen, Jiyuan and Peng, Minyi and Lam, Kwok-Yan and Yuan, Xingliang and Liu, Xiaoning},
  journal={ACM Computing Surveys},
  volume={57},
  number={1},
  pages={1--38},
  year={2024},
  publisher={ACM New York, NY}
}

@article{liu2024survey2,
  title={A Survey on Machine Unlearning: Techniques and New Emerged Privacy Risks},
  author={Liu, Hengzhu and Xiong, Ping and Zhu, Tianqing and Yu, Philip S},
  journal={arXiv preprint arXiv:2406.06186},
  year={2024}
}

@article{kingma2014adam,
  title={Adam: A method for stochastic optimization},
  author={Kingma, Diederik P},
  journal={arXiv preprint arXiv:1412.6980},
  year={2014}
}

@article{xing2024efuf,
  title={EFUF: Efficient Fine-grained Unlearning Framework for Mitigating Hallucinations in Multimodal Large Language Models},
  author={Xing, Shangyu and Zhao, Fei and Wu, Zhen and An, Tuo and Chen, Weihao and Li, Chunhui and Zhang, Jianbing and Dai, Xinyu},
  journal={arXiv preprint arXiv:2402.09801},
  year={2024}
}

@inproceedings{liu2023differentially,
  title={Differentially private sparse mapping for privacy-preserving cross domain recommendation},
  author={Liu, Weiming and Zheng, Xiaolin and Chen, Chaochao and Hu, Mengling and Liao, Xinting and Wang, Fan and Tan, Yanchao and Meng, Dan and Wang, Jun},
  booktitle={Proceedings of the 31st ACM International Conference on Multimedia},
  pages={6243--6252},
  year={2023}
}

@inproceedings{liu2024reducing,
  title={Reducing Item Discrepancy via Differentially Private Robust Embedding Alignment for Privacy-Preserving Cross Domain Recommendation},
  author={Liu, Weiming and Zheng, Xiaolin and Chen, Chaochao and Xu, Jiahe and Liao, Xinting and Wang, Fan and Tan, Yanchao and Ong, Yew-Soon},
  booktitle={Forty-first International Conference on Machine Learning},
  year={2024}
}

@inproceedings{su2024revisit,
  title={Revisit targeted model poisoning on federated recommendation: Optimize via multi-objective transport},
  author={Su, Jiajie and Chen, Chaochao and Liu, Weiming and Lin, Zibin and Shen, Shuheng and Wang, Weiqiang and Zheng, Xiaolin},
  booktitle={Proceedings of the 47th international acm sigir conference on research and development in information retrieval},
  pages={1722--1732},
  year={2024}
}

@inproceedings{su2023personalized,
  title={Personalized behavior-aware transformer for multi-behavior sequential recommendation},
  author={Su, Jiajie and Chen, Chaochao and Lin, Zibin and Li, Xi and Liu, Weiming and Zheng, Xiaolin},
  booktitle={Proceedings of the 31st ACM International Conference on Multimedia},
  pages={6321--6331},
  year={2023}
}

@inproceedings{lin2024data,
  title={Data-efficient Fine-tuning for LLM-based Recommendation},
  author={Lin, Xinyu and Wang, Wenjie and Li, Yongqi and Yang, Shuo and Feng, Fuli and Wei, Yinwei and Chua, Tat-Seng},
  booktitle={Proceedings of the 47th International ACM SIGIR Conference on Research and Development in Information Retrieval},
  pages={365--374},
  year={2024}
}

@inproceedings{acharya2023llm,
  title={Llm based generation of item-description for recommendation system},
  author={Acharya, Arkadeep and Singh, Brijraj and Onoe, Naoyuki},
  booktitle={Proceedings of the 17th ACM Conference on Recommender Systems},
  pages={1204--1207},
  year={2023}
}

@inproceedings{jia2018attriguard,
  title={$\{$AttriGuard$\}$: A practical defense against attribute inference attacks via adversarial machine learning},
  author={Jia, Jinyuan and Gong, Neil Zhenqiang},
  booktitle={27th USENIX Security Symposium (USENIX Security 18)},
  pages={513--529},
  year={2018}
}

@article{hu2022membership,
  title={Membership inference attacks on machine learning: A survey},
  author={Hu, Hongsheng and Salcic, Zoran and Sun, Lichao and Dobbie, Gillian and Yu, Philip S and Zhang, Xuyun},
  journal={ACM Computing Surveys (CSUR)},
  volume={54},
  number={11s},
  pages={1--37},
  year={2022},
  publisher={ACM New York, NY}
}

@inproceedings{ijcai2024p0238,
  title     = {Enhancing Dual-Target Cross-Domain Recommendation with Federated Privacy-Preserving Learning},
  author    = {Lin, Zhenghong and Huang, Wei and Zhang, Hengyu and Xu, Jiayu and Liu, Weiming and Liao, Xinting and Wang, Fan and Wang, Shiping and Tan, Yanchao},
  booktitle = {Proceedings of the Thirty-Third International Joint Conference on
               Artificial Intelligence, {IJCAI-24}},
  publisher = {International Joint Conferences on Artificial Intelligence Organization},
  pages     = {2153--2161},
  year      = {2024},
  month     = {8},
  note      = {Main Track},
}

@inproceedings{tan20224sdrug,
  title={4sdrug: Symptom-based set-to-set small and safe drug recommendation},
  author={Tan, Yanchao and Kong, Chengjun and Yu, Leisheng and Li, Pan and Chen, Chaochao and Zheng, Xiaolin and Hertzberg, Vicki S and Yang, Carl},
  booktitle={Proceedings of the 28th ACM SIGKDD Conference on Knowledge Discovery and Data Mining},
  pages={3970--3980},
  year={2022}
}

@inproceedings{koh2017understanding,
  title={Understanding black-box predictions via influence functions},
  author={Koh, Pang Wei and Liang, Percy},
  booktitle={International Conference on Machine Learning},
  pages={1885--1894},
  year={2017},
  organization={PMLR}
}

@article{koh2019accuracy,
  title={On the Accuracy of Influence Functions for Measuring Group Effects},
  author={Koh, Pang Wei W and Ang, Kai-Siang and Teo, Hubert and Liang, Percy S},
  journal={Advances in Neural Information Processing Systems},
  volume={32},
  pages={5254--5264},
  year={2019}
}

@inproceedings{basu2020second,
  title={On Second-Order Group Influence Functions for Black-Box Predictions},
  author={Basu, Samyadeep and You, Xuchen and Feizi, Soheil},
  booktitle={International Conference on Machine Learning},
  pages={715--724},
  year={2020},
  organization={PMLR}
}

@article{melchiorre2021investigating,
  title={Investigating gender fairness of recommendation algorithms in the music domain},
  author={Melchiorre, Alessandro B and Rekabsaz, Navid and Parada-Cabaleiro, Emilia and Brandl, Stefan and Lesota, Oleg and Schedl, Markus},
  journal={Information Processing \& Management},
  volume={58},
  number={5},
  pages={102666},
  year={2021},
  publisher={Elsevier}
}

@inproceedings{schedl2022lfm,
  title={LFM-2b: A dataset of enriched music listening events for recommender systems research and fairness analysis},
  author={Schedl, Markus and Brandl, Stefan and Lesota, Oleg and Parada-Cabaleiro, Emilia and Penz, David and Rekabsaz, Navid},
  booktitle={Proceedings of the 2022 Conference on Human Information Interaction and Retrieval},
  pages={337--341},
  year={2022}
}

@article{asghar2016yelp,
  title={Yelp dataset challenge: Review rating prediction},
  author={Asghar, Nabiha},
  journal={arXiv preprint arXiv:1605.05362},
  year={2016}
}

@inproceedings{graves2021amnesiac,
  title={Amnesiac machine learning},
  author={Graves, Laura and Nagisetty, Vineel and Ganesh, Vijay},
  booktitle={Proceedings of the AAAI Conference on Artificial Intelligence},
  volume={35},
  number={13},
  pages={11516--11524},
  year={2021}
}

@article{sekhari2021remember,
  title={Remember what you want to forget: Algorithms for machine unlearning},
  author={Sekhari, Ayush and Acharya, Jayadev and Kamath, Gautam and Suresh, Ananda Theertha},
  journal={Advances in Neural Information Processing Systems},
  volume={34},
  year={2021}
}

@inproceedings{basu2021influence,
  title={Influence Functions in Deep Learning Are Fragile},
  author={Basu, S and Pope, P and Feizi, S},
  booktitle={International Conference on Learning Representations (ICLR)},
  year={2021}
}

@article{gupta2021adaptive,
  title={Adaptive machine unlearning},
  author={Gupta, Varun and Jung, Christopher and Neel, Seth and Roth, Aaron and Sharifi-Malvajerdi, Saeed and Waites, Chris},
  journal={Advances in Neural Information Processing Systems},
  volume={34},
  pages={16319--16330},
  year={2021}
}

@inproceedings{wu2020mind,
  title={Mind: A large-scale dataset for news recommendation},
  author={Wu, Fangzhao and Qiao, Ying and Chen, Jiun-Hung and Wu, Chuhan and Qi, Tao and Lian, Jianxun and Liu, Danyang and Xie, Xing and Gao, Jianfeng and Wu, Winnie and others},
  booktitle={Proceedings of the 58th annual meeting of the association for computational linguistics},
  pages={3597--3606},
  year={2020}
}

@inproceedings{benjamin2018measuring,
  title={Measuring and regularizing networks in function space},
  author={Benjamin, Ari and Rolnick, David and Kording, Konrad},
  booktitle={International Conference on Learning Representations},
  year={2018}
}

@inproceedings{he2016fast,
  title={Fast matrix factorization for online recommendation with implicit feedback},
  author={He, Xiangnan and Zhang, Hanwang and Kan, Min-Yen and Chua, Tat-Seng},
  booktitle={Proceedings of the 39th International ACM SIGIR conference on Research and Development in Information Retrieval},
  pages={549--558},
  year={2016}
}

@inproceedings{resnick1994grouplens,
  title={Grouplens: An open architecture for collaborative filtering of netnews},
  author={Resnick, Paul and Iacovou, Neophytos and Suchak, Mitesh and Bergstrom, Peter and Riedl, John},
  booktitle={Proceedings of the 1994 ACM conference on Computer supported cooperative work},
  pages={175--186},
  year={1994}
}

@article{ludewig2021empirical,
  title={Empirical analysis of session-based recommendation algorithms: a comparison of neural and non-neural approaches},
  author={Ludewig, Malte and Mauro, Noemi and Latifi, Sara and Jannach, Dietmar},
  journal={User Modeling and User-Adapted Interaction},
  volume={31},
  number={1},
  pages={149--181},
  year={2021},
  publisher={Springer}
}

@inproceedings{ariannezhad2023personalized,
  title={A personalized neighborhood-based model for within-basket recommendation in grocery shopping},
  author={Ariannezhad, Mozhdeh and Li, Ming and Schelter, Sebastian and De Rijke, Maarten},
  booktitle={Proceedings of the Sixteenth ACM International Conference on Web Search and Data Mining},
  pages={87--95},
  year={2023}
}

@inproceedings{le2023constraint,
  title={A Constraint-based Recommender System via RDF Knowledge Graphs},
  author={Le, Ngoc Luyen and Abel, Marie-H{\'e}l{\`e}ne and Gouspillou, Philippe},
  booktitle={2023 26th International Conference on Computer Supported Cooperative Work in Design (CSCWD)},
  pages={849--854},
  year={2023},
  organization={IEEE}
}

@inproceedings{liu2021federaser,
  title={Federaser: Enabling efficient client-level data removal from federated learning models},
  author={Liu, Gaoyang and Ma, Xiaoqiang and Yang, Yang and Wang, Chen and Liu, Jiangchuan},
  booktitle={2021 IEEE/ACM 29th international symposium on quality of service (IWQOS)},
  pages={1--10},
  year={2021},
  organization={IEEE}
}

@article{quadrana2018sequence,
  title={Sequence-aware recommender systems},
  author={Quadrana, Massimo and Cremonesi, Paolo and Jannach, Dietmar},
  journal={ACM computing surveys (CSUR)},
  volume={51},
  number={4},
  pages={1--36},
  year={2018},
  publisher={ACM New York, NY, USA}
}

@inproceedings{wu2019session,
  title={Session-based recommendation with graph neural networks},
  author={Wu, Shu and Tang, Yuyuan and Zhu, Yanqiao and Wang, Liang and Xie, Xing and Tan, Tieniu},
  booktitle={Proceedings of the AAAI conference on artificial intelligence},
  volume={33},
  number={01},
  pages={346--353},
  year={2019}
}

@article{wang2025towards,
  title={Towards efficient and effective unlearning of large language models for recommendation},
  author={Wang, Hangyu and Lin, Jianghao and Chen, Bo and Yang, Yang and Tang, Ruiming and Zhang, Weinan and Yu, Yong},
  journal={Frontiers of Computer Science},
  volume={19},
  number={3},
  pages={193327},
  year={2025},
  publisher={Springer}
}

@inproceedings{chundawat2023can,
  title={Can bad teaching induce forgetting? unlearning in deep networks using an incompetent teacher},
  author={Chundawat, Vikram S and Tarun, Ayush K and Mandal, Murari and Kankanhalli, Mohan},
  booktitle={Proceedings of the AAAI Conference on Artificial Intelligence},
  volume={37},
  number={6},
  pages={7210--7217},
  year={2023}
}

@inproceedings{he2015trirank,
  title={Trirank: Review-aware explainable recommendation by modeling aspects},
  author={He, Xiangnan and Chen, Tao and Kan, Min-Yen and Chen, Xiao},
  booktitle={Proceedings of the 24th ACM International Conference on Information and Knowledge Management},
  pages={1661--1670},
  year={2015}
}

@inproceedings{liu2022backdoor,
  title={Backdoor defense with machine unlearning},
  author={Liu, Yang and Fan, Mingyuan and Chen, Cen and Liu, Ximeng and Ma, Zhuo and Wang, Li and Ma, Jianfeng},
  booktitle={IEEE INFOCOM 2022-IEEE conference on computer communications},
  pages={280--289},
  year={2022},
  organization={IEEE}
}

@inproceedings{liu2024backdoor,
  title={Backdoor attacks via machine unlearning},
  author={Liu, Zihao and Wang, Tianhao and Huai, Mengdi and Miao, Chenglin},
  booktitle={Proceedings of the AAAI Conference on Artificial Intelligence},
  volume={38},
  number={13},
  pages={14115--14123},
  year={2024}
}

@inproceedings{jia2021proof,
  title={Proof-of-learning: Definitions and practice},
  author={Jia, Hengrui and Yaghini, Mohammad and Choquette-Choo, Christopher A and Dullerud, Natalie and Thudi, Anvith and Chandrasekaran, Varun and Papernot, Nicolas},
  booktitle={2021 IEEE Symposium on Security and Privacy (SP)},
  pages={1039--1056},
  year={2021},
  organization={IEEE}
}

@inproceedings{thudi2022necessity,
  title={On the necessity of auditable algorithmic definitions for machine unlearning},
  author={Thudi, Anvith and Jia, Hengrui and Shumailov, Ilia and Papernot, Nicolas},
  booktitle={31st USENIX Security Symposium (USENIX Security 22)},
  pages={4007--4022},
  year={2022}
}

@article{weng2024proof,
  title={Proof of unlearning: Definitions and instantiation},
  author={Weng, Jiasi and Yao, Shenglong and Du, Yuefeng and Huang, Junjie and Weng, Jian and Wang, Cong},
  journal={IEEE Transactions on Information Forensics and Security},
  year={2024},
  publisher={IEEE}
}

@inproceedings{chen2021machine,
  title={When machine unlearning jeopardizes privacy},
  author={Chen, Min and Zhang, Zhikun and Wang, Tianhao and Backes, Michael and Humbert, Mathias and Zhang, Yang},
  booktitle={Proceedings of the 2021 ACM SIGSAC conference on computer and communications security},
  pages={896--911},
  year={2021}
}

@article{carlini2022privacy,
  title={The privacy onion effect: Memorization is relative},
  author={Carlini, Nicholas and Jagielski, Matthew and Zhang, Chiyuan and Papernot, Nicolas and Terzis, Andreas and Tramer, Florian},
  journal={Advances in Neural Information Processing Systems},
  volume={35},
  pages={13263--13276},
  year={2022}
}

@article{dangefficient,
  title={Efficient and Adaptive Recommendation Unlearning: A Guided Filtering Framework to Erase Outdated Preferences},
  author={Dang, Yizhou and Liu, Yuting and Yang, Enneng and Guo, Guibing and Jiang, Linying and Zhao, Jianzhe and Wang, Xingwei},
  journal={ACM Transactions on Information Systems},
  publisher={ACM New York, NY}
}

@inproceedings{feng2025plug,
  title={Plug and Play: Enabling Pluggable Attribute Unlearning in Recommender Systems},
  author={Feng, Xiaohua and Li, Yuyuan and Yu, Fengyuan and Zhang, Li and Chen, Chaochao and Zheng, Xiaolin},
  booktitle={Proceedings of the ACM Web Conference 2025},
  year={2025}
}

@inproceedings{yu2025lego,
  title={LEGO: A Lightweight and Efficient Multiple-Attribute Unlearning Framework for Recommender Systems},
  author={Yu, Fengyuan Yu and Li, Yuyuan Li and Feng, Xiaohua and Fang, Junjie and Wang, Tao and Chen, Chaochao},
  booktitle={Proceedings of the 33rd ACM International Conference on Multimedia},
  year={2025}
}

@inproceedings{escobedo2024simultaneous,
  title={Simultaneous unlearning of multiple protected user attributes from variational autoencoder recommenders using adversarial training},
  author={Escobedo, Gustavo and Ganh{\"o}r, Christian and Brandl, Stefan and Augstein, Mirjam and Schedl, Markus},
  booktitle={International Workshop on Algorithmic Bias in Search and Recommendation},
  pages={91--102},
  year={2024},
  organization={Springer}
}

@article{shu2024rah,
  title={Rah! recsys--assistant--human: A human-centered recommendation framework with llm agents},
  author={Shu, Yubo and Zhang, Haonan and Gu, Hansu and Zhang, Peng and Lu, Tun and Li, Dongsheng and Gu, Ning},
  journal={IEEE Transactions on Computational Social Systems},
  volume={11},
  number={5},
  pages={6759--6770},
  year={2024},
  publisher={IEEE}
}

@inproceedings{liu2025filtering,
  title={Filtering Discomforting Recommendations with Large Language Models},
  author={Liu, Jiahao and Shao, Yiyang and Zhang, Peng and Li, Dongsheng and Gu, Hansu and Chen, Chao and Du, Longzhi and Lu, Tun and Gu, Ning},
  booktitle={Proceedings of the ACM on Web Conference 2025},
  pages={3639--3650},
  year={2025}
}

@article{schelter2024snarcase,
  title={Snarcase-Regain Control over Your Predictions with Low-Latency Machine Unlearning},
  author={Schelter, Sebastian and Grafberger, Stefan and de Rijke, Maarten},
  journal={Proceedings of the VLDB Endowment},
  volume={17},
  number={12},
  pages={4273--4276},
  year={2024},
  publisher={VLDB Endowment}
}


\begin{IEEEbiography}[{\includegraphics[width=1in,height=1.25in,clip,keepaspectratio]{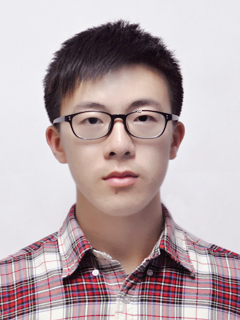}}]{Yuyuan Li} obtained his PhD degree in computer science from Zhejiang University, China, in 2023. He is currently an Associate Professor at Hangzhou Dianzi University. His research interests mainly focus on trustworthy machine learning. He has published over 30 papers in peer-reviewed journals and conferences, including Cell Patterns, NeurIPS, ICML, and ICLR.   

\end{IEEEbiography}

\begin{IEEEbiography}[{\includegraphics[width=1in,height=1.25in,clip,keepaspectratio]{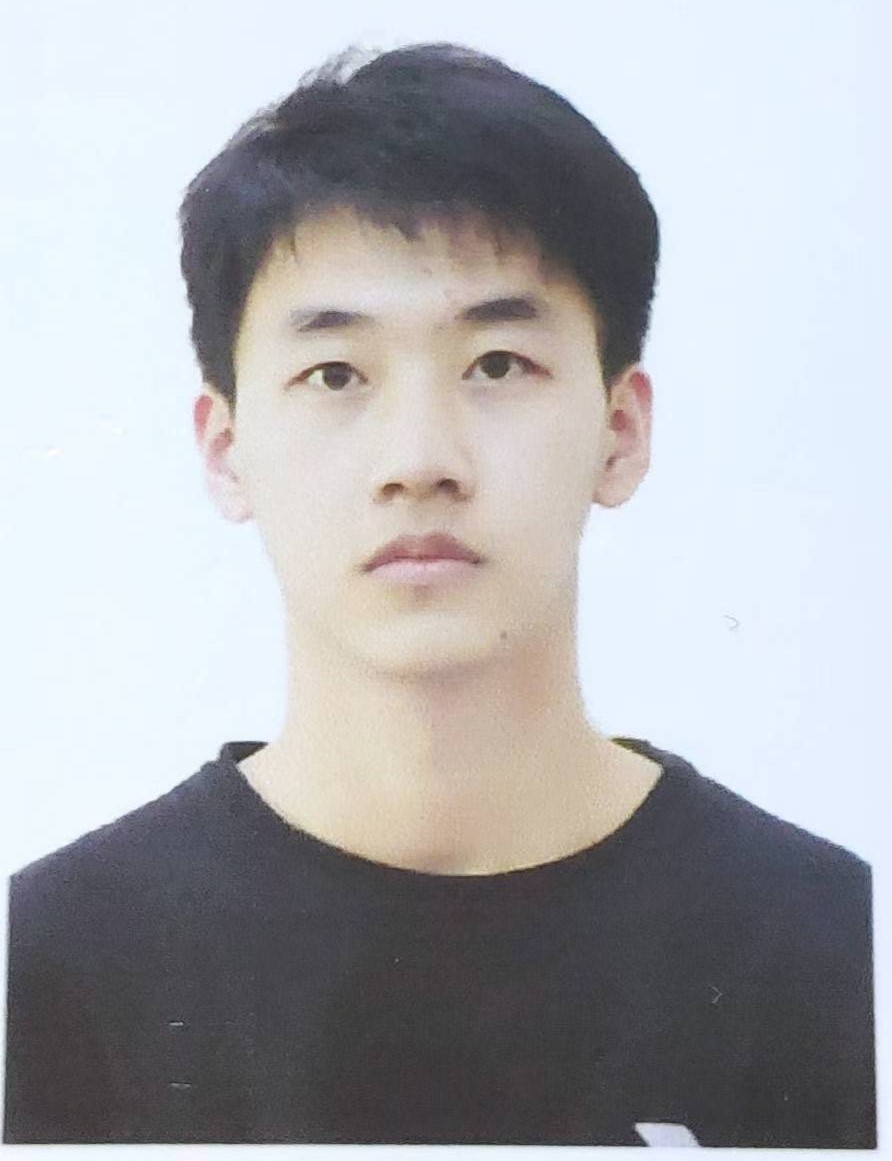}}]{Xiaohua Feng}
is currently pursuing a Ph.D. degree at the College of Computer Science and Technology, Zhejiang University, China. His undergraduate degree is from Wuhan University of Technology 2022. His research interests mainly focus on trustworthy machine learning. He has published papers in peer reviewed conferences and journals, including ICLR, WWW, EMNLP, and Cell Patterns.
\end{IEEEbiography}

\begin{IEEEbiography}[{\includegraphics[width=1in,height=1.25in,clip,keepaspectratio]{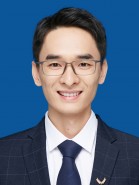}}]{Chaochao Chen} (Senior Member, IEEE) obtained his PhD degree in computer science from Zhejiang University, China, in 2016, and he was a visiting scholar in University of Illinois at Urbana-Champaign, during 2014-2015. He is currently a Distinguished Research Fellow at Zhejiang University. Before that, he was a Staff Algorithm Engineer at Ant Group. His research mainly focuses on recommender system, privacy preserving machine learning, and graph machine learning. He has published more than 100 papers in peer reviewed journals and conferences.
\end{IEEEbiography}

\begin{IEEEbiography}[{\includegraphics[width=1in,height=1.25in,clip,keepaspectratio]{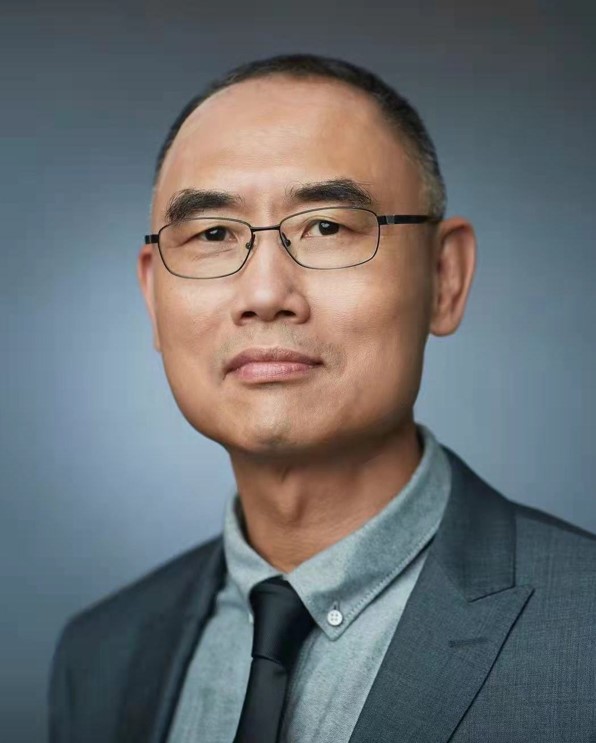}}]{Qiang Yang} (Fellow, IEEE) received the graduation degree from Peking University and PhD degree from the Department of Computer Science, the University of Maryland, College Park, in 1989 and had been a faculty member, the University of Waterloo between
1989 and 1995. He is a fellow of Canadian Academy of Engineering (CAE) and Royal Society of Canada (RSC), former Chief Artificial Intelligence Officer of WeBank, a former chair professor of Computer Science and Engineering Department at Hong Kong University of Science and Technology (HKUST). He is currently a Chair Professor and Director of Hong Kong Polytechnic University, Academy for AI and a Chair Professor at Hong Kong University of Science and Technology (Guangzhou).  He is the Conference Chair of AAAI-21, the Honorary vice president of Chinese Association for Artificial Intelligence(CAAI), the president of Hong Kong Society of Artificial Intelligence and Robotics(HKSAIR), the President of Investment Technology League (ITL) and Open Islands Privacy-Computing Open source Community. He is a fellow of AAAI, ACM, CAAI, IAPR, and AAAS. He was the Founding Editor in Chief of the ACM Transactions on Intelligent Systems and Technology (ACM TIST) and the Founding Editor in Chief of IEEE Transactions on Big Data (IEEE TBD). He received the ACM SIGKDD Distinguished Service Award, in 2017 and the Wu Wenjun outstanding contribution award of artificial intelligence, in 2019. His research interests are artificial intelligence, machine learning, data mining, and planning. He had been the Founding Director of the Huawei’s Noah’s Ark Research Lab between 2012 and 2015, the Founding Director of HKUST’s Big Data Institute, the Founder of 4Paradigm and the President of IJCAI (2017-2019). His latest books are Transfer Learning, Federated Learning, Privacy-preserving Computing, and Practicing Federated Learning.
\end{IEEEbiography}

\end{document}